\documentclass{aa}
\usepackage{times}
\usepackage{graphicx}
\usepackage{longtable}
\usepackage{psfig}
\usepackage{amsmath}
\usepackage{aas_macros}
\usepackage{amssymb}
\usepackage{natbib}
\bibpunct{(}{)}{;}{a}{,}{,}
\def\ngc#1{NGC$\,$#1}

\def\vmi{\hbox{\it V--I\/}}
\def\bmv{\hbox{\it B--V\/}}

\def\Deg{${}^\circ$\llap{.}}
\def\Min{${}^{\prime}$\llap{.}}
\def\Sec{${}^{\prime\prime}$\llap{.}}
\def\deg{${}^\circ$}
\def\min{${}^{\prime}$}
\def\sec{${}^{\prime\prime}$}
\begin{document}
   \title{ Parameter properties and stellar population of the old open cluster
   \ngc3960\thanks{Based on observations made with the European Southern Observatory
     telescopes obtained from the ESO/ST-ECF Science Archive Facility.}}

   \author{L. Prisinzano \inst{1},  G. Micela \inst{2}, S. Sciortino
   \inst{2}, F. Favata \inst{3}}

   \offprints{loredana@astropa.unipa.it }

   \institute{
Dipartimento di Scienze Fisiche ed Astronomiche, Universit\`a
di Palermo, Piazza del Parlamento 1, I-90134 Palermo Italy
\and INAF - Osservatorio Astronomico di Palermo, Piazza del Parlamento 1, 90134
Palermo Italy  \and Astrophysics Division - Space Science Department of ESA,
ESTEC, Postbus 299, 2200 AG Noordwijk, The Netherlands }

\date{Received 3 November 2003; accepted 18 December 2003}

   \abstract{We present a  $BV\!I$ photometric and  astrometric catalogue  of
the open cluster \ngc 3960, down to  limiting magnitude $V\sim22$, obtained
from observations taken with  the Wide Field Imager camera at the MPG/ESO 2.2\,m
Telescope at La Silla.  The photometry of all the stars detected in our field
of view has been used to estimate a map of  the strong   differential reddening
affecting this area.  Our results indicate that,  within the region where the
cluster dominates, the $E(V-I)$ values range from 0.21  up to 0.78,  with 
$E(V-I)=0.36$ ($E(B-V)=0.29$) at the nominal cluster  centroid position; color
excesses $E(V-I)$   up to 1 mag have been   measured in the external regions of
the field of view where field stars dominate. The reddening corrected
color-magnitude diagram (CMD)  allows us to conclude that the cluster has an
age between 0.9 and 1.4 Gyr  and a distance modulus of $(V-M_V)_0=11.35$. 

In order to minimize  field star contamination, their number has been
statistically  subtracted  based on the surface  density map. The empirical
cluster main sequence  has been recovered in the $V$ vs.  $V-I$ and in the
$J$ vs. $J-K_S$ planes, using optical and infrared data, respectively. From
these empirical cluster main sequences, two samples of candidate cluster
members were derived in order to obtain the luminosity distributions as a
function of the $V$ and $J$ magnitudes. 
 The Luminosity Functions have been transformed
into the corresponding Mass Functions; for $M>1~M_\odot$, the two
distributions have been fitted with a power law of index
$\alpha_V=2.95\pm0.53$ and $\alpha_J=2.81\pm0.84$ in $V$ and in $J$,
respectively, while the Salpeter Mass Function in this notation has index
$\alpha=2.35$.

   }

\titlerunning{\ngc3960} \authorrunning{Prisinzano et al.}
\maketitle 
\section{Introduction} Old Galactic open clusters are stellar
systems which can survive  only if they have a large initial population and lie
far from dense molecular clouds \citep{berg80,fuen97}.  For this reason they
are less numerous than the young open clusters and are mainly found in the
external region of the Galaxy, where there is  a lower  likelihood of 
catastrophic encounters. 

Cluster stars   
have the same distance and chemical composition; for this reason 
stellar clusters are among 
the most appropriate objects to study the  
Initial Mass Function (IMF), that is one of the most crucial ingredients for 
dynamical evolution models and for galaxy formation and stellar
evolution models. Comparison of the IMF of old open clusters with the IMF
of nearby and young open clusters is fundamental 
to look for similarities or discrepancies of the star formation processes 
in the Galaxy disk, in different epochs and in different environments. 

Accurate determination of the open cluster IMF is, however, 
a difficult task, because
 these objects are in general poorly populated and, due to their location in the
 thin disk, are highly contaminated by  field stars seen along
  the same line of sight.
In the case of old open clusters, these difficulties are increased,  
as they are
very distant from the Sun and thus
often strongly obscured due to interstellar absorption.
High-quality photometry and precise extinction determinations are crucial to
obtain
a correct estimate of the IMF.
 
With an age of about 1.2 Gyr  and a small spatial concentration, \ngc3960  
is an interesting example of an old open cluster suitable for IMF studies. 
It is a low-latitude cluster, 
located at the celestial and Galactic coordinates
RA=11$^h$50$^m$54$^s$  and Dec=$-$55\deg42\min (J2000)
and $l=$294\Deg42 and $b=$+6\Deg17, respectively,
about 1850 pc from the Sun.
  
Physical parameters of \ngc3960 were estimated by \citet{jane81}, through
 photoelectric {\em BV} and  intermediate-band David Dunlop Observatory
(DDO) photometry. One blue and one visual photographs  were obtained in order
to extend the magnitude  sequences to fainter limits. From the 
{\em BV} photometry, the author concludes that the cluster is 
$(0.5-1.0)\times10^9$ yr old, while from the DDO photometry for several 
cluster giants and using 
 the \citet{jane77} method, he derives the reddening $E(B-V)=0.29\pm0.02$.
  A distance modulus $(V-M_V)_0=11.1\pm0.2$ and
  metallicity [Fe/H]=$-0.30$ have also been estimated by the same author. 
 [Fe/H]=$-0.34\pm0.08$  was subsequently derived by \citet{frie93}
from a spectroscopic study of 7 giant stars of \ngc3960.

Because of its low-latitude position, in the direction of the Carina spiral feature
\citep{mill72},
  effects of differential reddening are
expected across the region of the sky including the cluster members. 
\citet{feit84} and \citet{hart86}
indicate the presence in their catalogues of two  dark nebulae falling in the
field of \ngc3960. More recently, \citet{dutr02} gave   values 
$E(B-V)=0.51$ and $E(B-V)=0.59$ for  
 these two overlapping 
dark nebulae, respectively, derived from  a 100 $\mu$m dust thermal emission map. 
These regions are located about 13 arcmin from the cluster center, where the 
$E(B-V)$ value is about 0.29 \citep{jane81}.  This non negligible 
difference in the $E(B-V)$  value, therefore, suggests the existence of variable
 extinction across the field of \ngc3960.
 
 In this work, using a procedure similar to that used by \citet{piot99a}
 and \citet{vonb01}, 
 we attempt to construct an extinction map of our field of view, in
 order to correct the photometric catalogue of candidate cluster members
 for differential reddening effects.
  
 In Section  \ref{photoastro}, we describe the observations and the data reduction
 procedure used to obtain the photometric/astrometric catalogue; in Section  
 \ref{colmag} we present the CMD and the method used to obtain the extinction map
 and the reddening corrected catalogue. In Section  \ref{param} we describe the
 method adopted to derive
  the cluster parameters, while in Section  \ref{removing} we  describe
 the statistical procedure used to recover the empirical cluster main sequence 
 from optical and infrared data and the photometric selection of candidate cluster
 members. In Section  \ref{lfmf}, we present the Luminosity Functions obtained from
 optical and infrared data and the corresponding Mass Functions. Finally, we
 summarize and discuss our results in Section  \ref{summary}.

\section{Cluster {\em  BVI} Photometry and Astrometry}
\label{photoastro}
\subsection{Observations  and Data Reduction \label{datared3960}}
\ngc3960 is an open cluster observed in the 
Pre-FLAMES Survey (PFS), carried out as part of the ESO Imaging Survey (EIS),
aimed to provide {\em  BVI}   imaging data
for use in connection with FLAMES, a multi-fiber spectrograph instrument on the
ESO VLT UT2 (KUEYEN) Telescope. 
The observations for this survey were
collected using the Wide Field Imager (WFI) camera mounted 
at the Cassegrain focus of the MPG/ESO 2.2\,m
Telescope at La Silla (Chile). This instrument consists of
a $4\times2$  mosaic of 
$2$k$\times 4$k CCD detectors with narrow inter-chip gaps of width 23\Sec8 
and 14\Sec3 along   right ascension and declination, respectively, 
yielding a filling factor of 95.9\%.  
With a pixel size  of 0.238\sec, each chip covers a field of view of
8\Min12 $\times$ 16\Min25, while the  
full field of view is 34\min $\times$ 33\min.

\begin{figure*}[!htb]
\includegraphics[width=15cm]{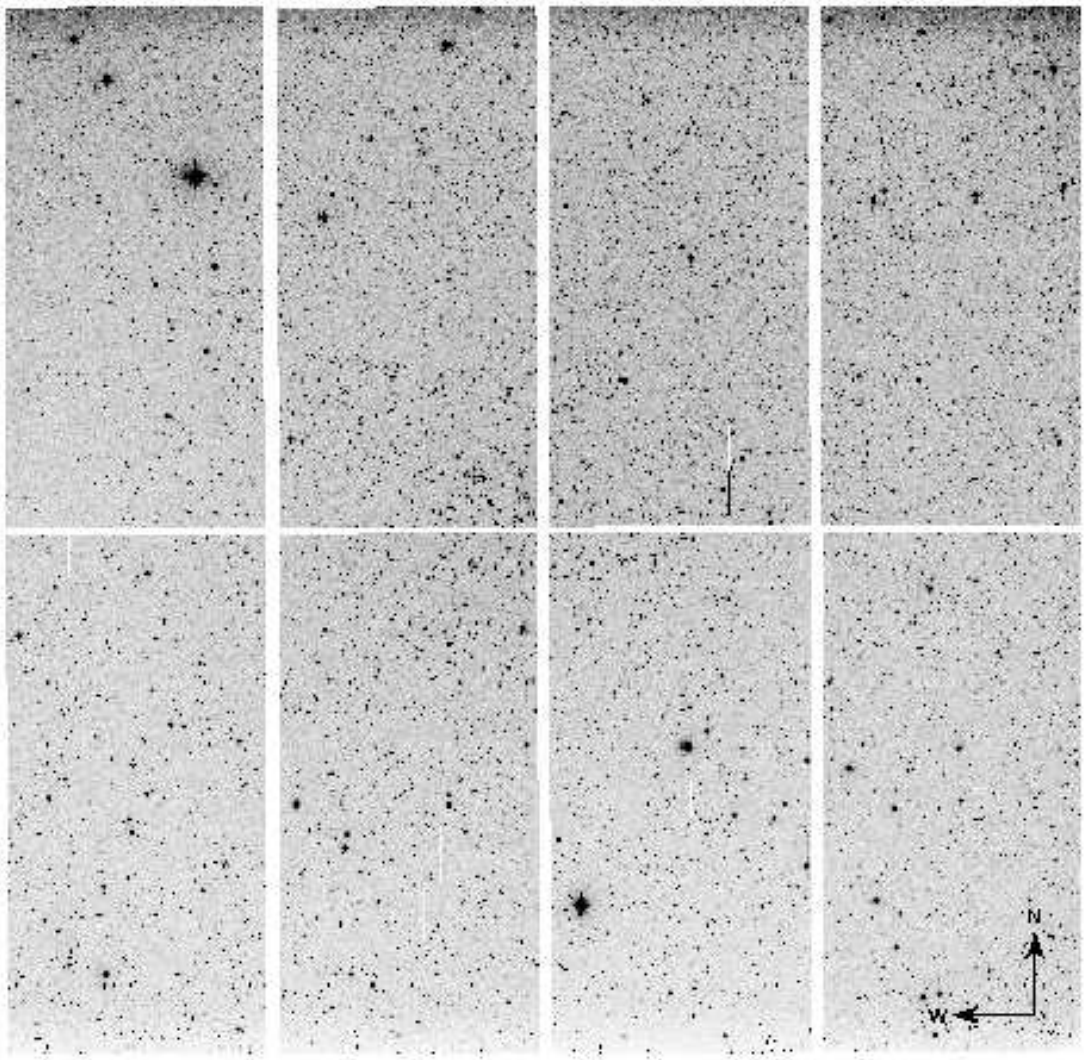}
\caption{A deep WFI image of the region around \ngc 3960 taken in the 
$B$ filter. The field center is RA=$11^h50^m51^{s}$\llap{.}6  and
Dec=$-55$\deg42\min07\sec (J2000). 
Each chip is 8\Min12 $\times$ 16\Min25, while the  
full field of view is 34\min $\times$ 33\min.
The cluster is approximately located in the central part of the field of view
where it is barely  visible within a radius of about 7 arcmin.
Strong effects of differential reddening cause a non-uniform spatial
distribution of the field stars in 
the external regions  of the field of view.}
\label{oc21deepB}
\end{figure*}

According to the PFS observing strategy, the images were obtained in the
 {\em  BVI} pass-bands; for each filter,  one short exposure of
30 seconds, useful to avoid saturating bright objects, and two deep 
exposures, of 4 minutes each,  have been taken. In order to cover the  
inter-chip gaps, these
observations were dithered by 30\sec~ both in right ascension and
declination. Table \ref{obs3960} gives the log-book of the observations and
 Figure  \ref{oc21deepB} shows a deep image of our field of view in the $V$ band.
The cluster is approximately located in the central part of the field of view
where it is barely  visible within a radius of about 7 arcmin.
Strong effects of differential reddening cause a non-uniform spatial
distribution of the field stars in 
the external regions  of the field of view. 
 In addition to the science target exposures, a set of technical frames 
for the instrumental calibration were obtained during the same night. 
\begin{table*}
\centering
\tabcolsep 0.1truecm
\caption {Log-book of the  observations.}
\vspace{0.5cm}
\begin{tabular}{ ccccccc} 
\hline
\hline
\multicolumn{1}{c}{Target}&
\multicolumn{1}{c}{RA (J2000)}&
\multicolumn{1}{c}{Dec (J2000)}&
\multicolumn{1}{c}{Night}&
\multicolumn{1}{c}{Filter}&  
\multicolumn{1}{c}{Exp. Time} &
\multicolumn{1}{c}{seeing} \\  
EIS name & (h m s)  & (d m s) & & &[s] & FWHM[\sec] \\
\hline\\
\ngc 3960 & 11 50 55.0 &$-$55 41 36 & 24--25 Feb 2000 & {\it B}&
$1\times30+2\times240$  
& 0.99--1.21\\
(OC21) & & &  & {\it V} & $1\times30+2\times240$   & 0.86--1.38 \\
& &&& {\it I} & $1\times30+2\times240$   & 0.81--1.19 \\

\hline
\hline
\end{tabular}
\label{obs3960}
\end{table*}

To complement the optical data, we have utilized $JHK_S$ infrared (IR) photometry 
from the   All-Sky Point Source Catalogue
of the Two Micron All Sky Survey (2MASS)\footnote{a joint project of 
the University of Massachusetts and the Infrared Processing and
Analysis Center/California Institute of Technology, funded by the National 
Aeronautics and Space Administration and the National Science Foundation}
\citep{carp01} available on the WEB\footnote{http://irsa.ipac.caltech.edu/}.

The instrumental calibration of the WFI CCD images has been 
performed using the {\tt mscred} package, a
mosaic specific task implemented as an IRAF package for the {\it NOAO
Mosaic Data Handling System} (MDHS). 
 
 The first stage of the data reduction process has been the 
removal of the instrumental signatures from each raw CCD image. 
First of all, we have subtracted the electronic bias using the overscan
region that has been subsequently removed.  
Exposure bias patterns have been
subtracted using a zero calibration mosaic obtained as the median of two 
bias frames. Finally, flat fielding for each filter
has been performed using a set of dome
flat fields combined into a median master flat.
 Due to some saturated points, we could not use the sky flat fields.  

Images in the $I$ band have required a special treatment because of the strong
effects of fringing. In order to remove this instrumental artifact we have 
subtracted
the fringing pattern provided by the MPG/ESO 2.2\,m Telescope 
team\footnote{available at  
http://www.ls.eso.org/lasilla/Telescopes/2p2T/E2p2M/WFI},
scaling it to each exposures.

Instrumental magnitudes were obtained for each chip of each mosaic image
by using the DAOPHOT II/ALLSTAR \citep{stet87} and ALLFRAME \citep{stet94}
photometric routines.
The point-spread functions (PSFs), determined  
for each chip and filter, take into account the variation in the PSF 
across the field of view of each WFI chip. 

 In order to obtain the aperture correction to the profile-fitting
 photometry, growth curves were determined. First, we have
selected the stars used to define the PSF model; next, all
other objects were removed from the frames and aperture photometry was
carried out at a variety of radii. DAOGROW \citep{stet90} was used to
derive growth curves and  COLLECT to 
calculate the "aperture correction" coefficient for each chip from 
the difference between PSF and aperture magnitudes of the selected stars.

Calibration to the Johnson-Cousin photometric system was performed by means
of a set of \citet{land92} standard fields (SA98, SA104, PG0918 and Ru 152), 
observed during the same night through the 8 chips.

In order to derive the transformation coefficients  to the standard system
we have used   equations of the form:
\begin{eqnarray}
 v &= V + A_0 + A_1 \times  X + A_2 \times (\bmv), \nonumber\\
 b &= B + B_0 + B_1 \times  X + B_2 \times (\bmv), \\
 i &= I + C_0 + C_1 \times  X + C_2 \times (\vmi). \nonumber   
\end{eqnarray}  
where  $v$, $b$ and $i$  are the standard star instrumental magnitudes,
corrected to the exposure time of 1 s; $X$ is the airmass and
$V$, $B$ and $I$ are the magnitudes in the standard system.   
We first attempted to derive all the coefficients 
(the zero points, the extinction and the color terms) for each chip, 
but by analyzing the
photometric residuals as a function of the time
through the night, 
we noticed that when the PG0918
standard field was observed,
the  night  was not photometric, as the mean residuals  systematically   
change with time. Since the problem does not affect the other science
observations,
we have discarded all the observations of the PG0918 field and we have 
only
used the Landolt fields SA98, SA104 and Rubin 152.
In order to improve the photometric calibration, 
the number of the standard stars in each of these fields has been increased using 
secondary standards defined in \citet{stet00}. Unfortunately, 
neither the Landolt's   nor the Stetson's catalogue  cover the
chips 4 and 5 of our field with a sufficient number of standards to
determine independent
coefficients for these chips. Neighboring chips have been used to calibrate 
science observations in these
chips, as described at the end of this section.
Imposing mean extinction coefficients (Stetson, private communication),
a zero point and a color term were determined separately for the chips  
where a sufficient number of standard observations were available. 
We have found that the color terms 
computed for the chips 2, 3, 6 and 7 are consistent with each other, with
variations $\lesssim 0.01$ in $B$, $\lesssim 0.02$ in $V$ and 
$\lesssim 0.05$ in $I$; for the  chips 1 and 8 we have found different 
values because they have been computed on  
a smaller number of standard stars (see Table
\ref{calib3960}). Considering the homogeneous characteristics of the instrument,
we have chosen to compute 
new zero points   fixing the color term to the
median value of the color terms of the chips 
for which  the number of measured standard stars was sufficiently large.  
 The resulting coefficients  
 are given in Table \ref{calib3960} together with the corresponding average standard
 errors. Points above the $3~\sigma$ level were not considered 
 for the calibration. We note that  	
 the variations among the photometric zero points of different chips 
 (Table \ref{calib3960}),
  are $\lesssim 0.11 $ in $V$, $\lesssim 0.14$ in $B$ and $\lesssim 0.08$ 
 in I, 
 in agreement with the results found in \citet{zocc03}.
\begin{table*}[!htb]
\centering
\caption {Coefficients of the transformation to the standard system for each filter
and for each chip. Col. 3 gives the number of standard stars used to derive
the zero points and the color terms (Col. 4 and 5), imposing the mean extinction 
coefficients of La Silla (Col. 6); finally, Col. 7 gives the corresponding average 
standard errors.}
\vspace{0.5cm}
\begin{tabular}{ccccccc} 
\hline
\hline
Filter& Chip $\#$& N. stars.& Zero Point& Color Term &Extinction& 
Av. St. Error\\
\hline
$V$   &1	 & 24	&$0.987 \pm 0.007$&    -0.097 &0.14	& 0.034\\
      &2	 &306	&$1.022 \pm 0.002$&    -0.097 &0.14	& 0.031\\
      &3	 &136	&$1.062 \pm 0.003$&    -0.097 &0.14	& 0.030\\
      &4	 &  4	&		  &    -0.097 &0.14	&     \\
      &5	 &  0	&		  &    -0.097 &0.14	&     \\
      &6	 & 51	&$1.095 \pm 0.005$&    -0.097 &0.14	& 0.033\\
      &7	 & 67	&$1.048 \pm 0.003$&    -0.097 &0.14	& 0.022\\
      &8	 & 11	&$0.982 \pm 0.007$&    -0.097 &0.14	& 0.024\\

\hline

$B$	&1	 & 25	&$0.668 \pm 0.012$&   0.249 &0.25 	& 0.053\\
 	&2	 &328	&$0.723 \pm 0.002$&   0.249 &0.25 	& 0.038\\
 	&3	 &148	&$0.772 \pm 0.002$&   0.249 &0.25 	& 0.028\\
	&4	 &  4	&		  &   0.249 &0.25 	&   \\
	&5	 &  0	&		  &   0.249 &0.25 	&   \\
 	&6	 & 49	&$0.810 \pm 0.008$&   0.249 &0.25 	& 0.054\\
	&7	 & 56	&$0.754 \pm 0.004$&   0.249 &0.25 	& 0.030\\
	&8	 & 11	&$0.697 \pm 0.019$&   0.249 &0.25 	& 0.062\\

\hline

$I$	&1	 & 22	&$1.989 \pm 0.012$&   0.130 &0.09      & 0.056\\
 	&2	 &248	&$2.042 \pm 0.002$&   0.130 &0.09      & 0.035\\
 	&3	 &123	&$2.063 \pm 0.003$&   0.130 &0.09      & 0.038\\
	&4	 &  6	&		  &   0.130 &0.09      &    \\
	&5	 &  0	&		  &   0.130 &0.09      &    \\
 	&6	 & 42	&$2.072 \pm 0.007$&   0.130 &0.09      & 0.047\\
	&7	 & 47	&$2.066 \pm 0.005$&   0.130 &0.09      & 0.033\\
	&8	 & 13	&$2.009 \pm 0.007$&   0.130 &0.09      & 0.026\\
 
\hline
\hline
\end{tabular}
\label{calib3960}
 \end{table*}

 As already mentioned, the \ngc3960 field observations
 were dithered both in
 right ascension and declination to cover the inter-chip  gaps. This means that
 contiguous chips have several stars in common. We have compared the 
 magnitudes of 
 these stars to verify consistency of our photometry among different chips.  
 The mean value and the root mean square of the residuals 
 for the stars with $V<20$ are given in Table. \ref{dmag}.
\begin{table*}[!htb]
\centering
\caption {The mean value and the root mean square of the $V$, $B$ and $I$
residuals of the common stars of contiguous chips. }
\vspace{0.5cm}
\begin{tabular}{ccccccc} 
\hline
\hline
Chips& $<\Delta V>$ &rms & $<\Delta B> $& rms& $<\Delta I>$ &rms\\
 \hline
1-2 &  0.051 &  0.023&   0.043 &  0.025 &  0.054 &  0.047\\
1-8 & -0.006 &  0.027&   0.033 &  0.072 &  0.009 &  0.020\\
2-3 &  0.005 &  0.036&   0.004 &  0.055 & -0.018 &  0.031\\
2-7 &  0.021 &  0.070&   0.043 &  0.049 &  0.014 &  0.051\\
3-6 &  0.046 &  0.021&   0.035 &  0.036 &  0.034 &  0.017\\
6-7 & -0.032 &  0.029&  -0.071 &  0.055 & -0.023 &  0.026\\
7-8 & -0.065 &  0.039&  -0.015 &  0.043 & -0.043 &  0.027\\
\hline
\hline
\end{tabular}
\label{dmag}
 \end{table*} 
 Finally, the photometric zero points of the   chips 3 and 6 have been 
used to calibrate the stars in the chips 4 and 5, respectively.
 Corrections  $\lesssim 0.02$, $\lesssim 0.01$ and $\lesssim 0.02$ mag to 
 these photometric zero points 
 have been found, for the magnitudes $V$, $B$ and $I$, respectively, 
 using the photometric
comparison of the common stars in the overlapping regions.  

%
Completeness of our star list and accuracy of our photometry 
for each chip have been determined by  
tests with artificial stars; a total of about 1350 stars have been added
as described in \citet{pris03}.
Our data are $100\%$ complete above $V=20$ and
 more than $50\%$ complete  above $V\sim 21$, $I\sim 19.5$ and
$B\sim 22$. The data for the recovered artificial stars of each chip
have been sorted by the observed magnitude and divided
into 14 group of about 80 stars. For each group, the median observed magnitude
$\bar{V}$ and the external errors
have been computed as in \citet{stet88}. In Table \ref{photoer3960}, we report
the estimated errors for $V>16.5$. The values for $V<16.5$ are less than 0.001
mag.
\begin{table*}[!htb]
\centering
\tabcolsep 0.1truecm
\caption {External photometric errors estimated from the artificial star
experiments.}
\vspace{0.5cm}
\begin{tabular}{ccccc|ccccc} 
\hline
\hline
Chip & $\bar{V}$& $\sigma_{ext}(V)$& $\sigma_{ext}(B-V)$& $\sigma_{ext}(V-I)$&  
Chip &$\bar{V}$& $\sigma_{ext}(V)$& $\sigma_{ext}(B-V)$& $\sigma_{ext}(V-I)$\\
\hline
1& 16.721 &  0.003 &  0.005 &  0.004& 5& 16.873 &  0.003 &  0.006 &  0.003 \\
1& 17.617 &  0.006 &  0.009 &  0.007& 5& 18.024 &  0.007 &  0.013 &  0.007 \\
1& 18.663 &  0.010 &  0.021 &  0.013& 5& 19.128 &  0.016 &  0.037 &  0.018 \\
1& 19.746 &  0.027 &  0.047 &  0.032& 5& 20.111 &  0.033 &  0.062 &  0.043 \\
1& 20.878 &  0.065 &  0.150 &  0.083& 5& 21.456 &  0.113 &  0.213 &  0.105 \\
\hline				      
2& 16.738 &  0.003 &  0.006 &  0.004& 6& 16.911 &  0.003 &  0.006 &  0.004 \\
2& 17.916 &  0.006 &  0.009 &  0.004& 6& 18.049 &  0.006 &  0.013 &  0.006 \\
2& 18.856 &  0.015 &  0.021 &  0.010& 6& 19.051 &  0.015 &  0.034 &  0.018 \\
2& 19.989 &  0.028 &  0.062 &  0.027& 6& 20.079 &  0.040 &  0.085 &  0.043 \\
2& 21.185 &  0.094 &  0.188 &  0.093& 6& 21.459 &  0.110 &  0.254 &  0.116 \\
\hline				      
3& 16.755 &  0.003 &  0.006 &  0.003& 7& 16.995 &  0.003 &  0.007 &  0.003 \\
3& 18.064 &  0.007 &  0.015 &  0.009& 7& 18.084 &  0.009 &  0.018 &  0.010 \\
3& 19.313 &  0.019 &  0.033 &  0.019& 7& 19.144 &  0.016 &  0.037 &  0.013 \\
3& 20.321 &  0.034 &  0.086 &  0.032& 7& 20.155 &  0.035 &  0.070 &  0.039 \\
3& 21.318 &  0.093 &  0.227 &  0.126& 7& 21.441 &  0.120 &  0.246 &  0.095 \\
\hline				     
4& 16.980 &  0.003 &  0.004 &  0.003& 8& 16.804 &  0.003 &  0.005 &  0.003 \\
4& 17.910 &  0.006 &  0.013 &  0.007& 8& 17.897 &  0.007 &  0.013 &  0.007 \\
4& 18.942 &  0.016 &  0.034 &  0.012& 8& 18.983 &  0.013 &  0.036 &  0.013 \\
4& 19.987 &  0.031 &  0.087 &  0.033& 8& 20.088 &  0.033 &  0.072 &  0.038 \\
4& 21.277 &  0.074 &  0.231 &  0.065& 8& 21.283 &  0.095 &  0.170 &  0.081 \\
\hline
\hline
\end{tabular}
\label{photoer3960}
 \end{table*} 
\subsection{Astrometry}
\begin{figure}[!htb]
\includegraphics[width=7.5cm]{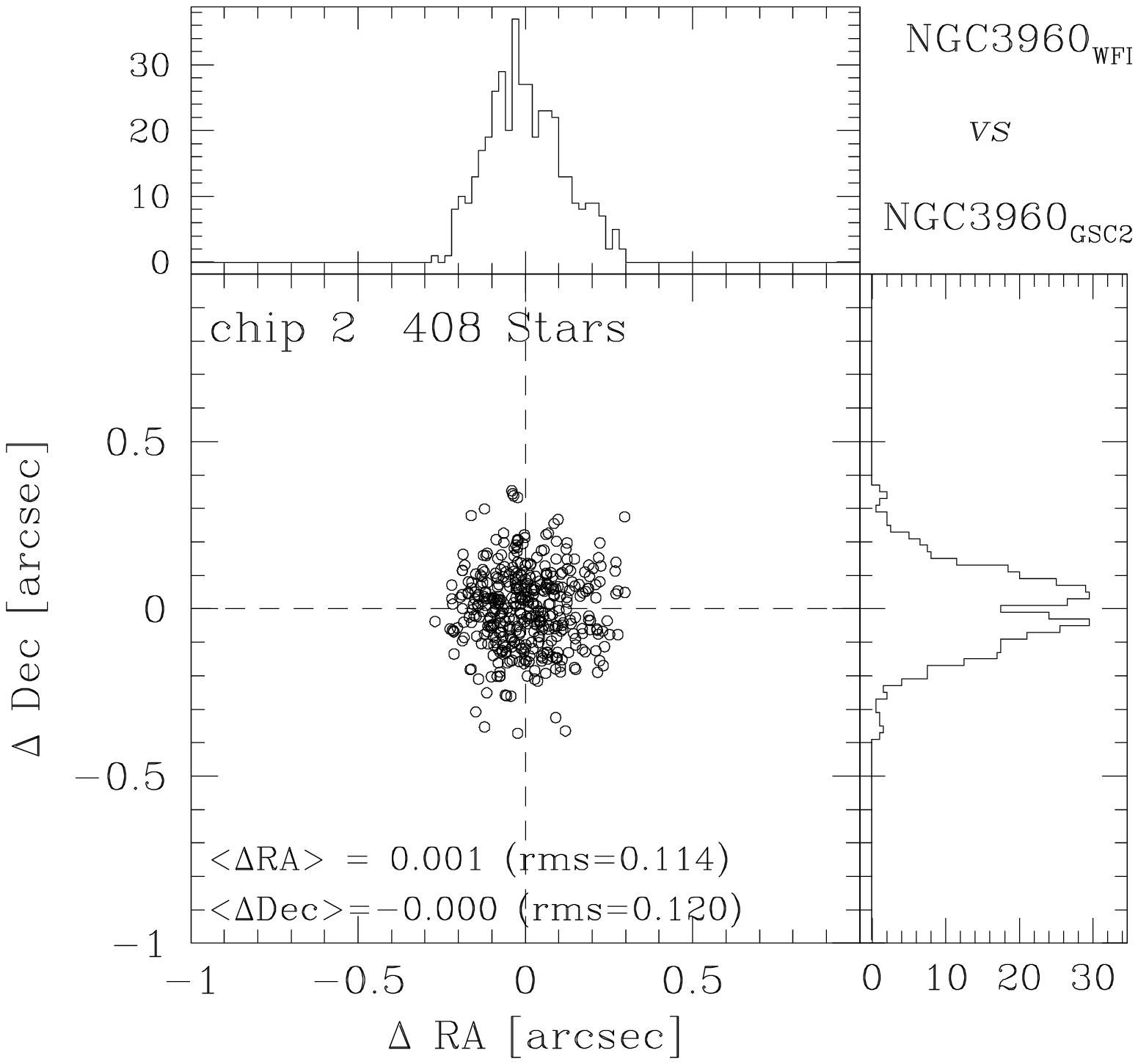}
\includegraphics[width=7.5cm]{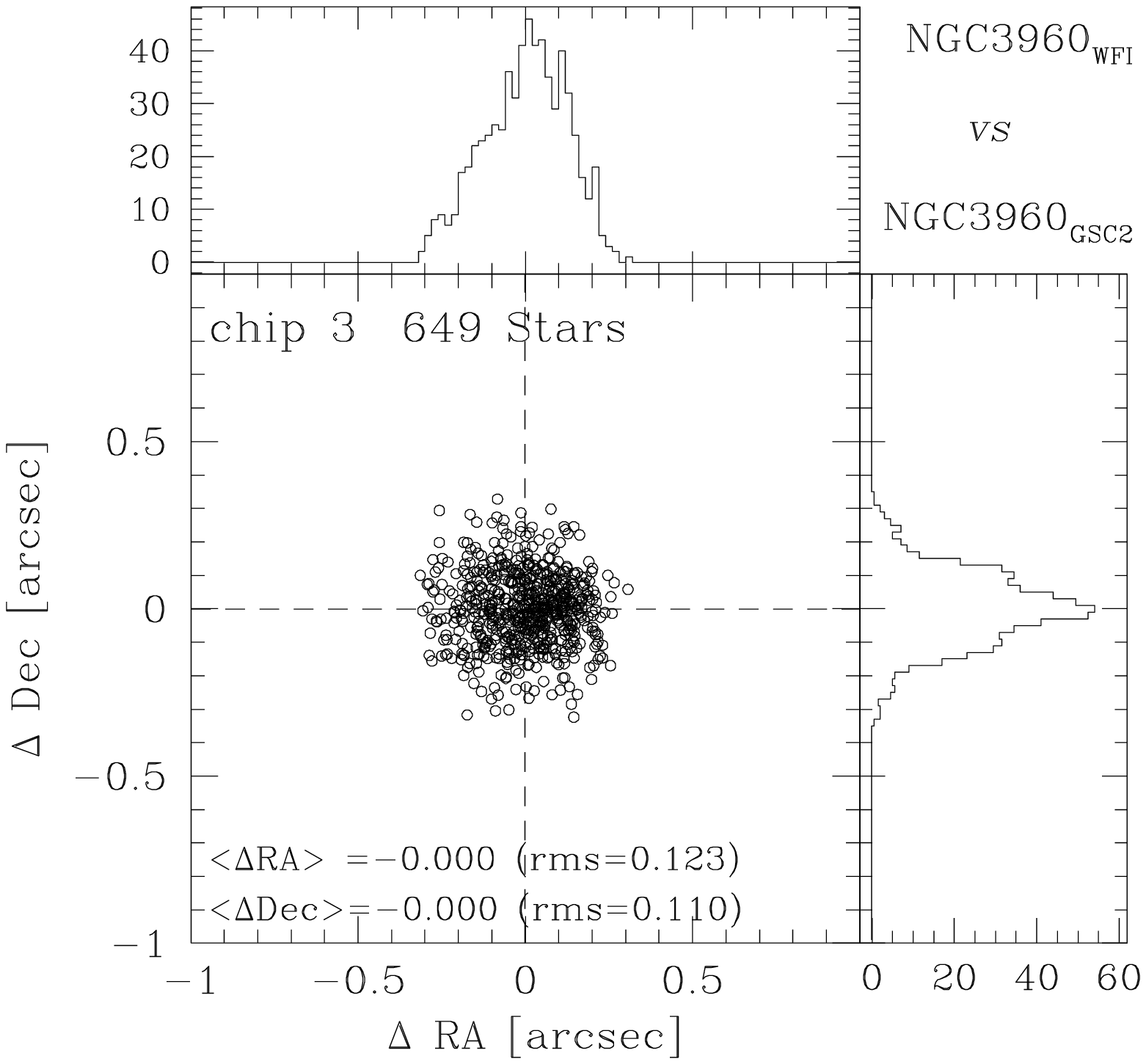}
\caption{Scatter plot of the  RA and Dec residuals of the transformation 
and distributions of $\Delta$RA and $\Delta$Dec  
obtained from the comparison of our astrometric solution and the GSC\,2.2 
reference catalogue. The vertical and horizontal dashed lines mark the mean
residuals in RA and Dec.
The results are shown only for the chips 2 ({\it left panel})
and 3 ({\it right panel}), where the
cluster is mainly concentrated. The number of   matched stars, the mean
values and the $1~\sigma$ rms  in arcsec are given on the figures.}
\label{resrade}
\end{figure}

In order to determine the astrometric solution 
we have considered   as reference catalogue  the Guide Star Catalogue, 
Version 2.2.01 (GSC\,2.2  STScI, 2001).
We have used the Aladin Sky Atlas to 
  plot  this catalogue on the finding chart of the open cluster \ngc 3960
  field. For each chip we have chosen three stars for which we have both the 
  celestial
  coordinates of the GSC\,2.2 catalogue and the pixel coordinates. 
  We have used these stars as reference for the IRAF task {\tt ccxymatch};
   imposing a matching tolerance of 0\Sec3,
    we have matched a total of 4224  stars. The IRAF task {\tt ccmap} has been used
    to fit a transformation between pixel coordinates and celestial coordinates
    after applying a $3~\sigma$ clipping to the data.
 In our case, 
 a tangent plate projection with distortion polynomials has been chosen
 for the sky projection geometry.
The scatter plot and the distributions of the RA and Dec residuals of 
the transformation are shown 
 in Figure  \ref{resrade} for the chips 2 and 3 where the cluster is mainly
 concentrated. We note that the residual distribution is highly 
 concentrated with a 
 Gaussian shape in both directions. The mean offsets and the rms of the distributions,
 given on each panel, show  that
 the final accuracy is always better than 0\Sec2.
\begin{figure}[!h]
\includegraphics[width=7.5cm]{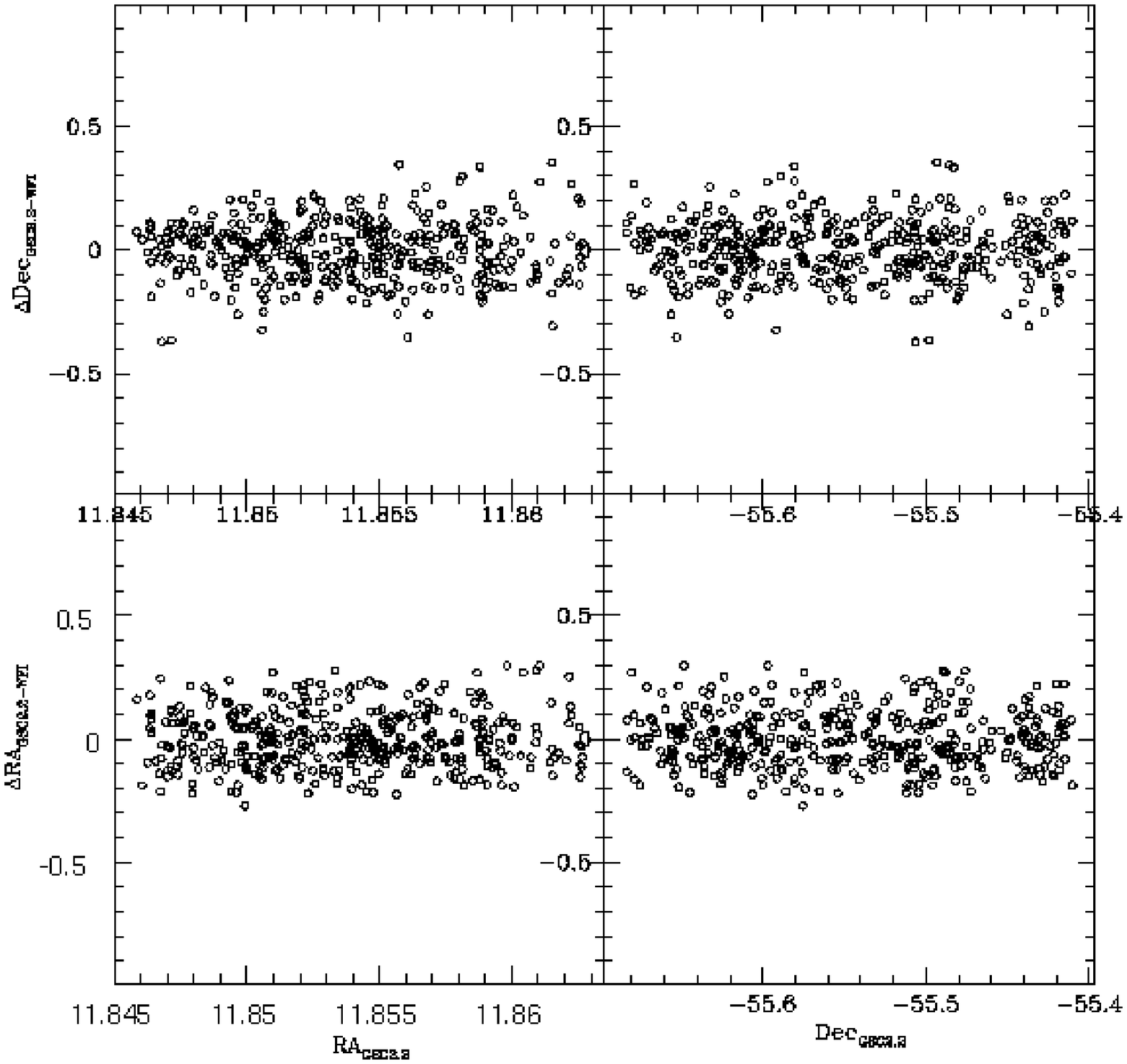}
\includegraphics[width=7.5cm]{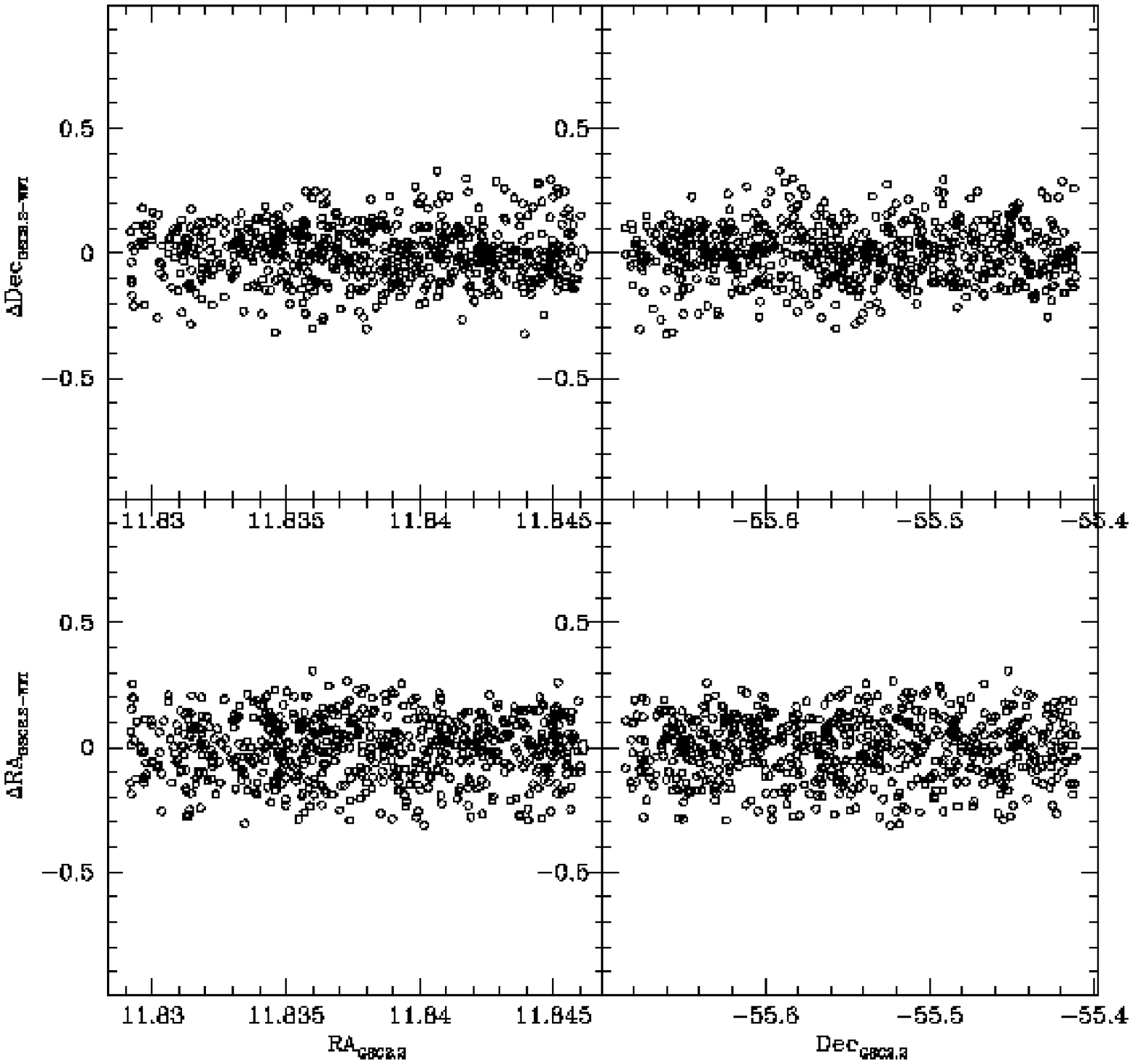}
\caption{Positional residuals, computed as in Figure  \ref{resrade},  
as a function of RA and Dec in the GSC\,2.2 reference catalogue. The results refer
to chips 2 ({\it left panel}) and 3 ({\it right panel}), 
where the cluster is mainly  concentrated.}
\label{respos}
\end{figure} 
 In order to investigate eventual systematic dependence from the position, we
have plotted the RA and Dec residuals as a function of RA and Dec.
 The results for
the chips 2 and 3 are shown
in Figure  \ref{respos} where no noticeable systematic effects are evident. 
Similar results have been found for the other chips.
%
\section{The Color-Magnitude Diagrams  }
\label{colmag}
\begin{figure*}[!ht]
\centerline{\psfig{file=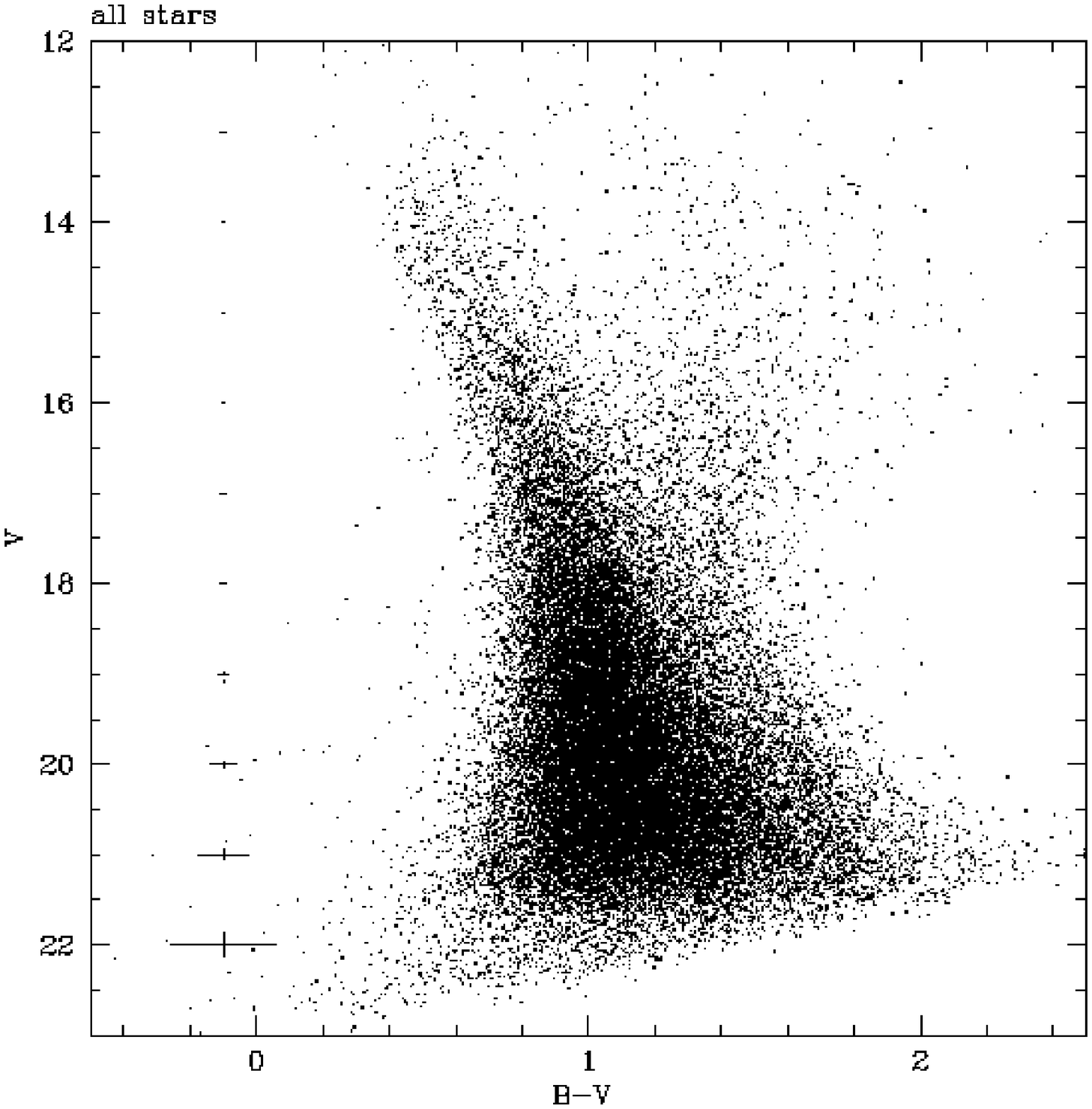,width=8cm,height=10cm} \psfig{figure=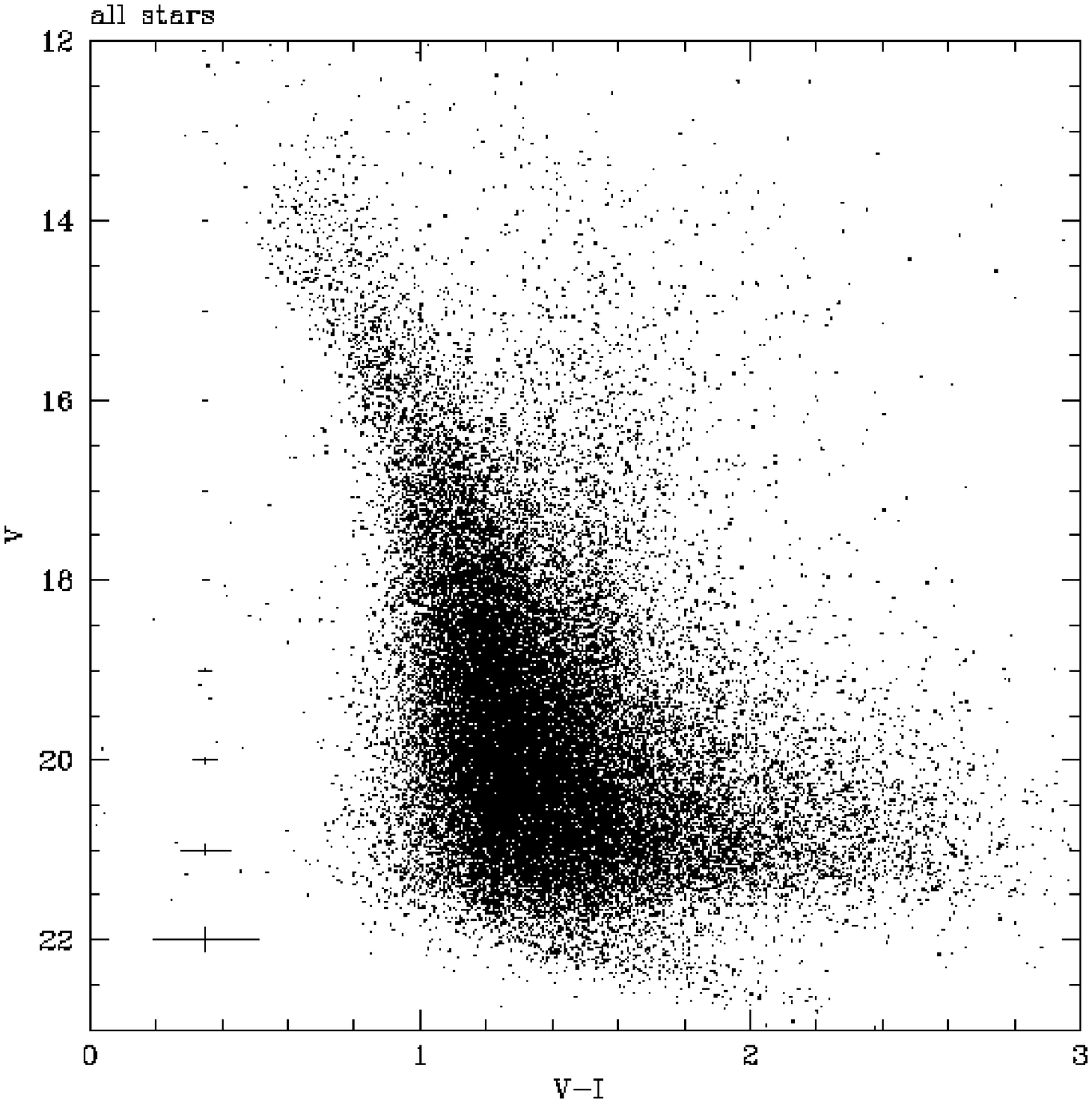,width=8cm,height=10cm}}
\centerline{\psfig{file=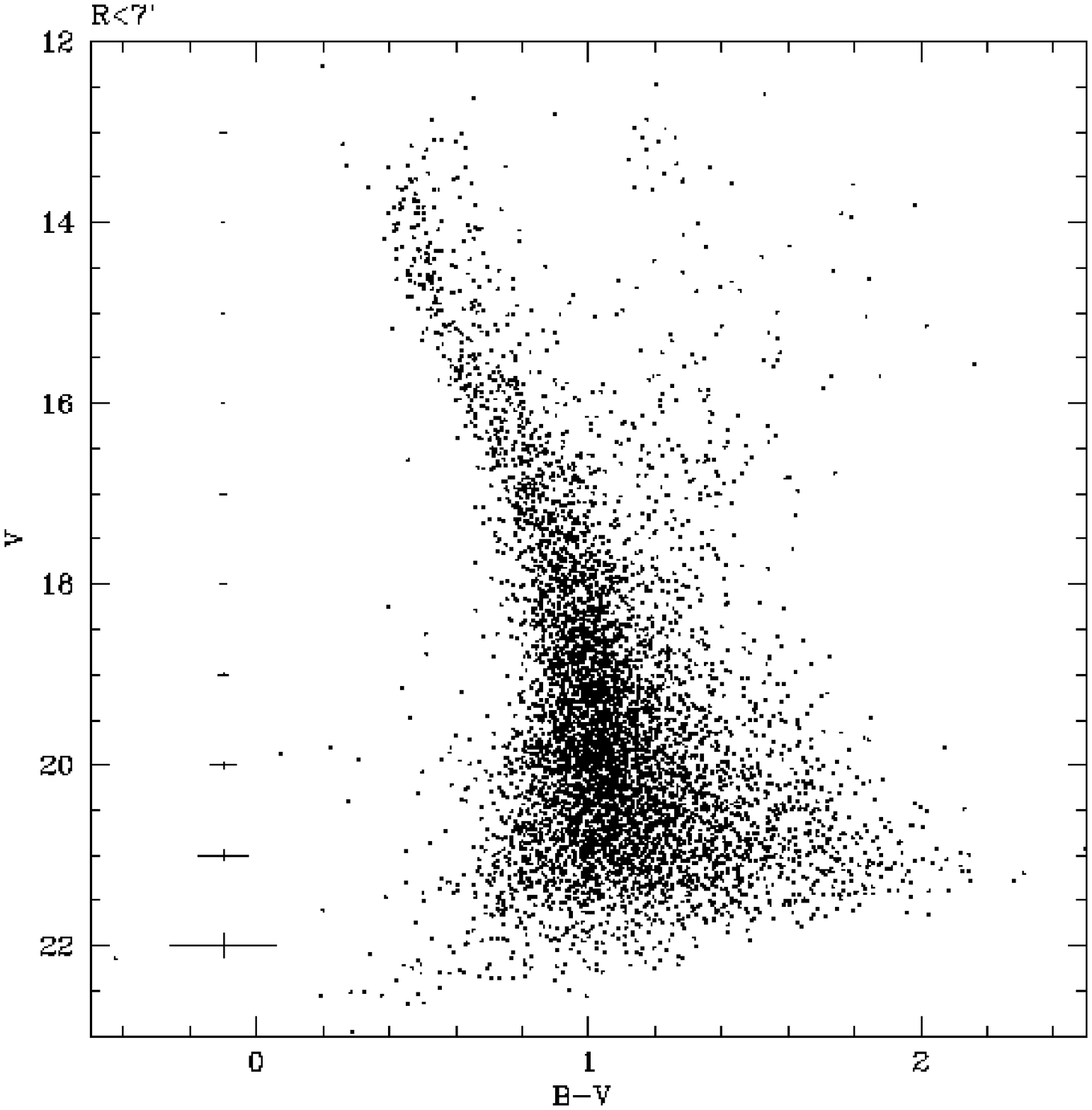,width=8cm,height=10cm} \psfig{figure=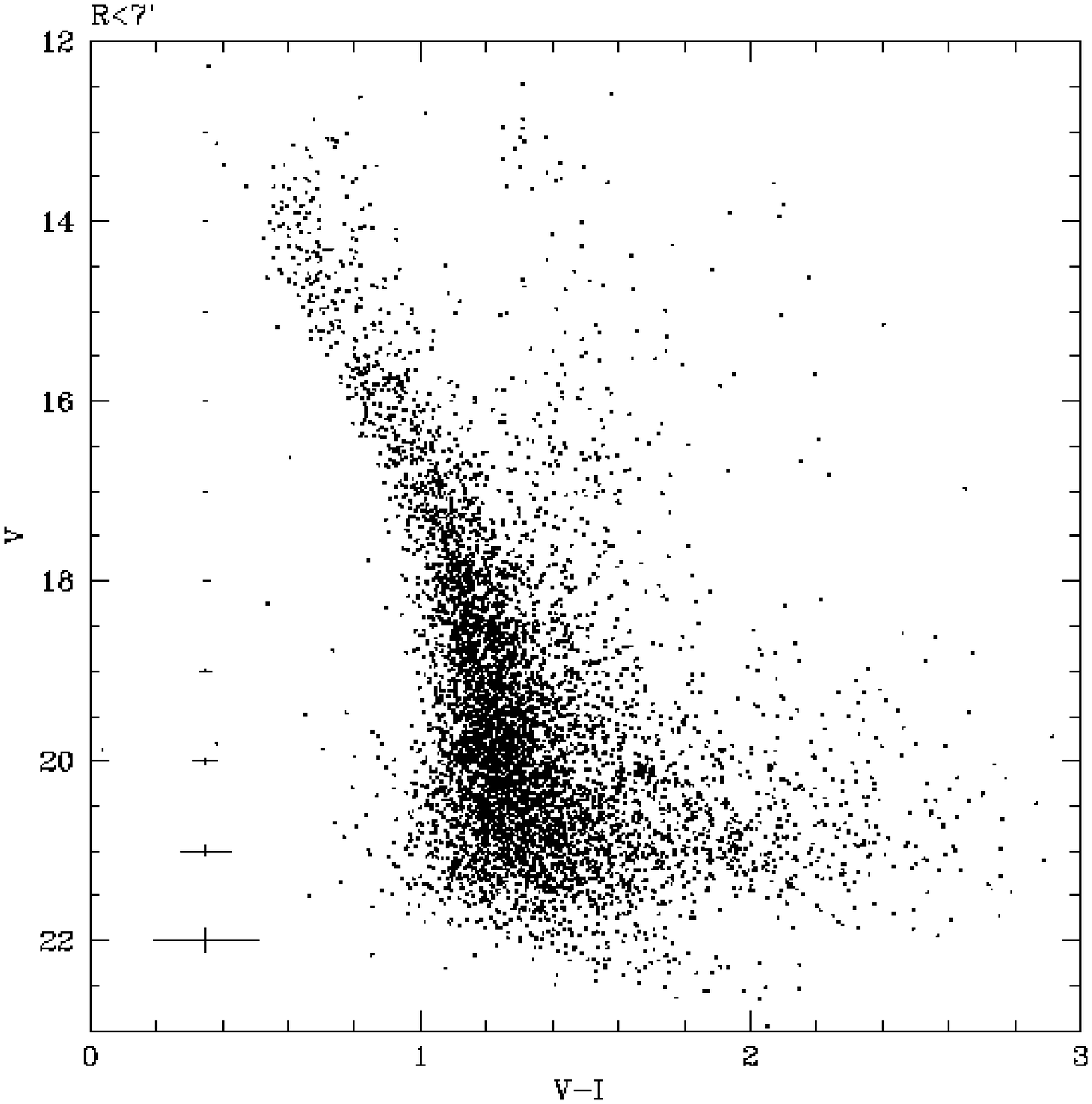,width=8cm,height=10cm}}
\caption{The top panels show the $V$ vs. $B-V$ and the  $V$ vs. $V-I$ 
color-magnitude diagrams of all the stars measured in our field of view, while
the bottom panels show the same diagrams of all the stars within the "{\it
cluster region}".}
\label{rawcmd}
\end{figure*}

\begin{figure}[!ht]
\includegraphics[width=9.0cm]{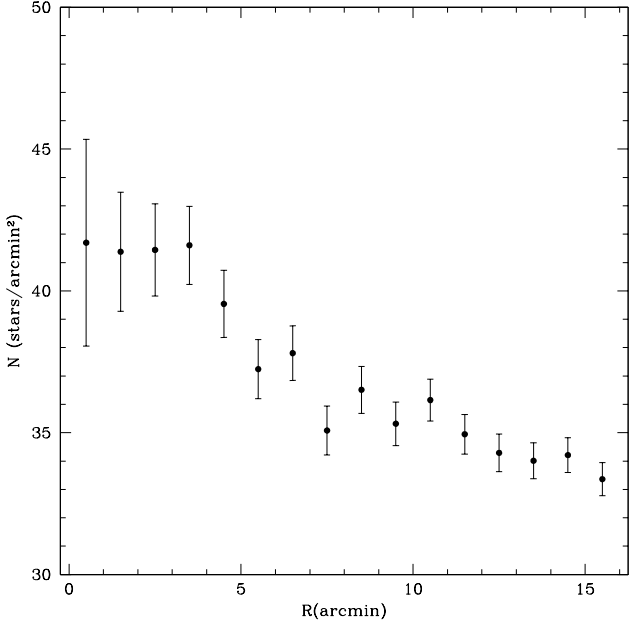}
\caption{Star counts as a function of the distance from the centroid
 for all stars detected in the
field of \ngc 3960.}
\label{radius}
\end{figure}

The color-magnitude diagrams of all the stars measured in our field of view
have been obtained using all sources with {\tt SHARP} parameter 
\citep{stet87} between
$-$0.8 and 0.8. This selection allows us to reject non-stellar objects
such as semi-resolved galaxies or blended double stars or cosmic rays.
The  $V$ vs. $B-V$ and the $V$ vs.  $V-I$ 
color-magnitude diagrams of the 39\,411 selected stars are shown in
Figure  \ref{rawcmd} (top panels). Due to the Galactic position
of the cluster,   the diagrams are highly contaminated by foreground and
background stars making it difficult to discriminate the cluster main sequence.
To facilitate the interpretation of  the color-magnitude diagrams,
we have considered the distribution of the stars as a function of   distance
from the cluster center.  
For this   we have first calculated the cluster centroid 
RA$_{\rm cen}=11^h50^m41^{s}$\llap{.}8 and 
Dec$_{\rm cen}=-55$\deg40\min37\Sec4 (J2000),
as the median value of the celestial positions of stars brighter than V=15.5. 
Therefore we have performed stars counts by using our data to obtain a
first order approximation
of the cluster size. We have derived the surface stellar density by performing
star counts in concentric rings around the cluster centroid and then dividing
them by their respective areas. Figure  \ref{radius} shows the resulting density
profile and the corresponding Poisson errors bars.
We note a flattening of the profile outside 7 arcmin.
 This density variation suggests that most of the cluster members are located
within 7 arcmin, whereas outside the Galactic disk population is dominant.
However, we consider this radius only an approximate value
 where the cluster dominates the field.
 In fact, as expected from dynamical evolution and mass segregation
effects, faint cluster stars can be located out of this radius.
We note, however, that our cluster size estimate is significantly
  greater than the value 2.75
arcmin given for the radius of this cluster by \citet{jane81}, where only
stars  brighter than $V=17$ were considered.
From here on we refer to the above mentioned region inside the circle
of 7 arcmin as the "{\it cluster region}" and to the remaining part of our field
of view as the "{\it field region}".

The  color-magnitude diagrams obtained  using all the stars within  
the cluster region  are shown in Figure  \ref{rawcmd} (bottom panels). 
Both the diagrams exhibit the cluster main sequence down to $V=17$ but 
the faint star main sequence is still
hidden by contaminating stars.
The Turn Off (TO) point is roughly located at $(V=14,~B-V=0.3)$, 
while the group of bright stars in the red part of the CMD
are clearly Red Giant Branch (RGB) stars of \ngc 3960.  

\subsection{The Differential Reddening Map}
\label{difredmap}
\begin{figure*}[!ht]
\centerline{\psfig{figure=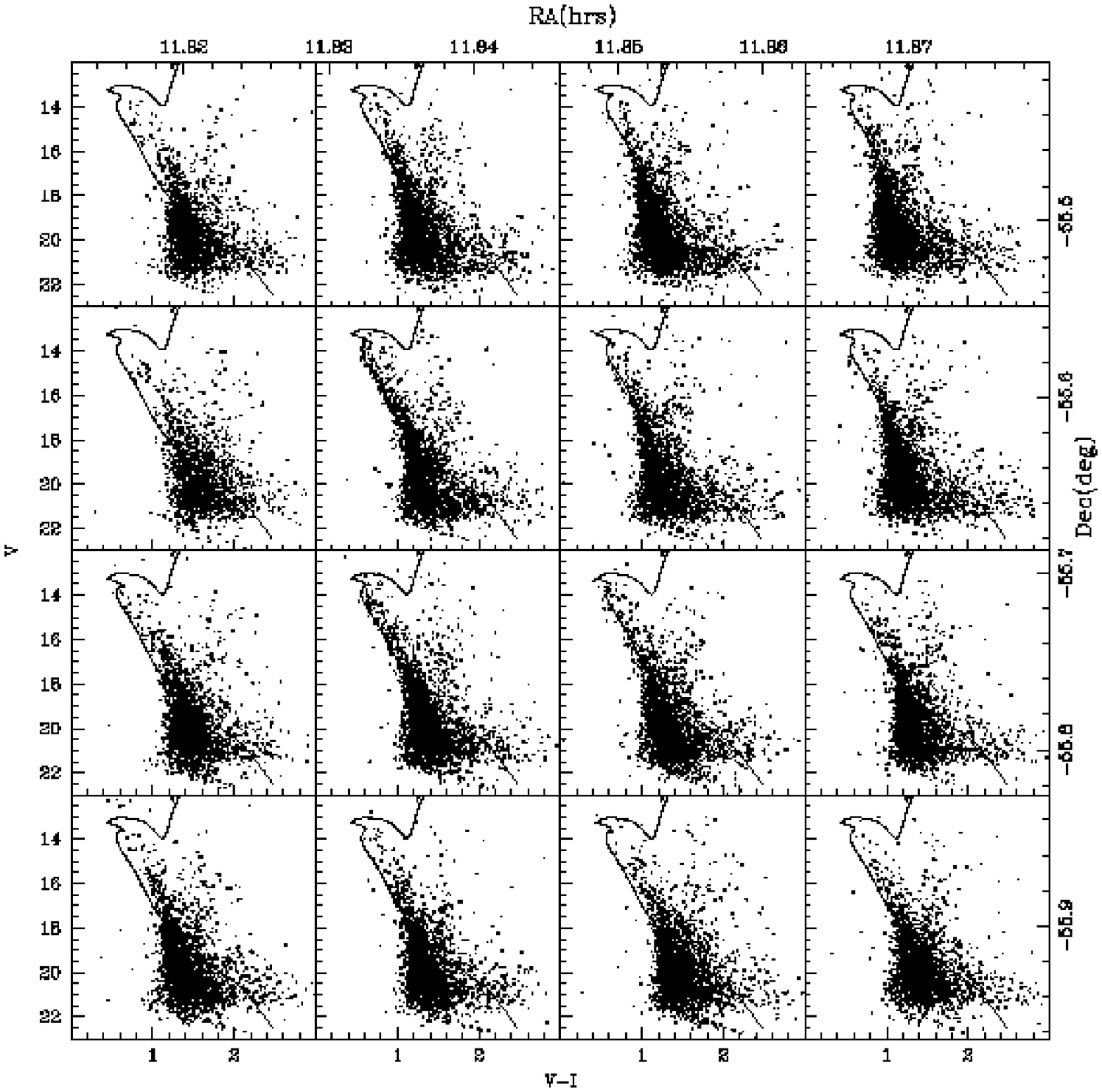,width=17cm,height=17cm}}
\caption{$V$ vs. $V-I$  color-magnitude diagrams
of 16 subregions of about 8\Min6$\times$8\Min3 
in the field of \ngc 3960. The solid line is a representative
theoretical isochrone of 1.1 Gyr and metallicity $Z=0.01$, calculated
 by  \citet{piet03}.}
\label{redvi}
\end{figure*}

The cluster main sequence in Figure  \ref{rawcmd} 
appears rather broad  in both   diagrams, with  higher dispersion in the
$V$ vs. $V-I$  CMD (lower-right panel).
Artificial-star tests indicate that our typical  photometric errors
for $V<17$ are $\sigma_{(B-V)} \sim 0.007$ and 
$\sigma_{(V-I)}\sim 0.004$; therefore,
after ruling out   instrumental causes and/or photometric errors
as possible   origin for this effect, we have
investigated a possible dependence of such broadening from the positions 
of the stars in our field of view. 
In Fig  \ref{redvi}
we show the  $V-I$ vs $V$ color-magnitude
diagrams of 16 subregions of about 8\Min6$\times$8\Min3 in the
 \ngc 3960 field. Celestial
coordinates corresponding to each subregion are given on the top and right 
axes, respectively. The solid line on the color-magnitude diagrams is the
isochrone of 1.1 Gyr,  metallicity $Z=0.01$ 
computed by the \citet{piet03} (see Section  \ref{param}). 

We note that the cluster main sequence is 
dominant in the  central regions,  while the number of  
stars in the upper main sequence decreases  in 
the external regions.
Nevertheless, a different appearance of the main sequence   as a function
of the position is clearly evident as regards  the theoretical model. 
This effect could be due to the fact that the star cluster is present only
in one small well defined region;
such a difference is however also evident looking at the
blue edge of the  field population, roughly located  
from 0.8 to 1.2 in $(V-I)$.  
We note that this shift does not depend from the chip to chip
photometric zero point offsets  because it  is present in 
different CMDs of stars falling in the same chip; in addition
differences  in these CMD are not consistent with  photometric
zero point residuals of adjacent chips 
found in Section  \ref{datared3960}. Thus, we conclude that 
the dependence of the appearance
on the spatial positions of the stars  is real and can be due to differential
reddening.    
 
This conclusion is consistent with several previous studies indicating
the presence of  dust clouds in the \ngc 3960 field. For instance,
a region of 0.116 square degrees, located at the  
RA=11$^h$49$^m$10$^s$, Dec=$-$55\deg31\min40\sec (J2000)
and ($l=$294\Deg15, $b=$6\Deg28),  and 
 falling in the north-west part of our field of view, has been included
in a catalogue of dark nebulae compiled by  \citet{feit84}. In this study, many
 fields of the ESO/SRC southern sky survey were examined looking for 
regions of the sky where the apparent surface density is reduced compared to
the surrounding regions. The  dark nebula falling in our field of view
was classified as a cloud with some structure and relatively well defined edges.
Another dark cloud, falling also
in the north-west part of our field of view, was identified by 
\citet{hart86} using the ESO/SERC Southern $J$ survey. 
The catalogue obtained
in this work includes a dark nebula  of minimum density
of size 5\min$\times$3\min  ~centered on RA=11$^h$49$^m$03$^s$ and 
Dec=-55\deg40\min22\sec (J2000)
($l=$294\Deg16, $b=$6\Deg13). The distances of these dark clouds are
 not known.

More recently, 
a value for the reddening of these regions
was given in the unified catalogue of dust clouds by \citet{dutr02}.
The reddening values were extracted from the all-sky reddening
$E(B-V)_{\rm FIR}$ map  of \citet{schl98},
based on a 100 $\mu$m dust thermal emission map with resolution of $\simeq6$\min.
 For each region, the authors compare the  reddening  in the nebula 
direction ($E(B-V)_{\rm cen}$)
with the background reddening ($E(B-V)_{\rm bck}$) computed
from the average of the measured values
in four background surrounding positions, used as reference. In our field
they found 
$E(B-V)_{\rm cen}=0.51$ and  $E(B-V)_{\rm bck}=0.38$, for the first dark nebula and
$E(B-V)_{\rm cen}=0.59$ and $E(B-V)_{\rm bck}=0.55$, for the second one.
These values are significantly different from the cluster reddening value
$E(B-V)=0.29$
estimated by \citet{jane81} as the average  of reddening measurements ranging
from $E(B-V)=0.20$ and $E(B-V)=0.34$ obtained for 6 giants in \ngc 3960.
Such  reddening estimates indicate that  strong gradients  of 
interstellar absorption are present in our field, confirming our conclusion on the
presence of differential reddening deduced by looking at the color-magnitude
diagrams.

In order to quantify such effects on the color-magnitude diagrams
we have created two reddening maps, one  for 
the     stars  located at the cluster distance, using the cluster main sequence,
and
a  reddening map for the field star population, affected by a further reddening
spread due to distance. Both maps are used to apply an average correction
 for differential
reddening to the photometric data.

As in \citet{piot99a},
the main sequence reddening map has been calculated as follows:
we divided  our field in 20 $\times$ 20 subregions of about  
1\Min7$\times$1\Min7  where we have obtained the CMDs. Such subregion size is a
good compromise to 
have a sufficient number of stars to identify the cluster main sequence,
at least within 7 arcmin from the cluster centroid, and 
high spatial resolution to map the differential reddening.
We have selected as fiducial region the subregion with coordinates 
RA =[$11^h50^m28^s$\llap{.}3, $11^h50^m45^s$\llap{.}6] and  
Dec =[-55\deg41\min13\Sec2, $-55$\deg38\min50\Sec6] (J2000)
where the 
cluster main sequence is well defined and where our estimated cluster centroid
is located; we have fitted with a spline the points of the CMD where the star 
density is larger, defining the  cluster main sequence;
for each star  with  $12<V<18.2$ and 
$0.4<(V-I)<2.4$, we have calculated the distance from the fitted sequence
along the reddening vector defined by the relation 
$A_V=3.1/1.25\times E(V-I)$, where $E(V-I)=1.25\times E(B-V)$ \citep{muna96}. The
relative reddening has been calculated as the median value of the 
$V-I$ component of these distances.

\begin{figure*}[!ht]
\centerline{\psfig{figure=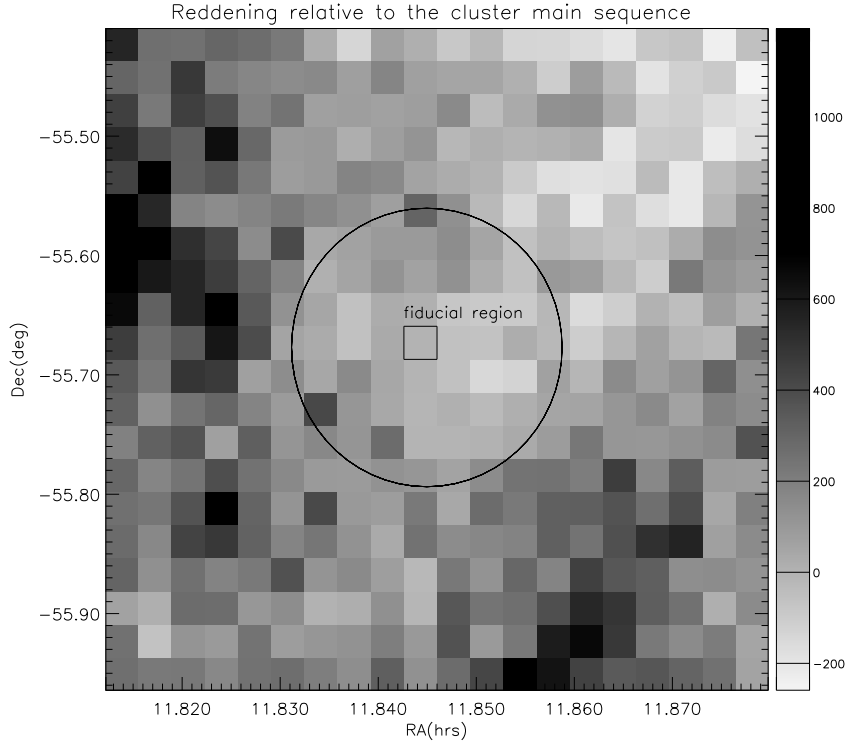,width=11.0cm,height=11.0cm}}
\caption{Greyscale reddening map of the \ngc3960 cluster members.
 In this figure,
the darker regions correspond to   higher relative reddening, calculated 
with respect to the fiducial cluster main sequence.  
The position of the fiducial region
and the cluster region are also indicated as the small box and the circle,
respectively.}
\label{redmapseq}
\end{figure*}

\begin{figure*}[!ht]
\centerline{\psfig{figure=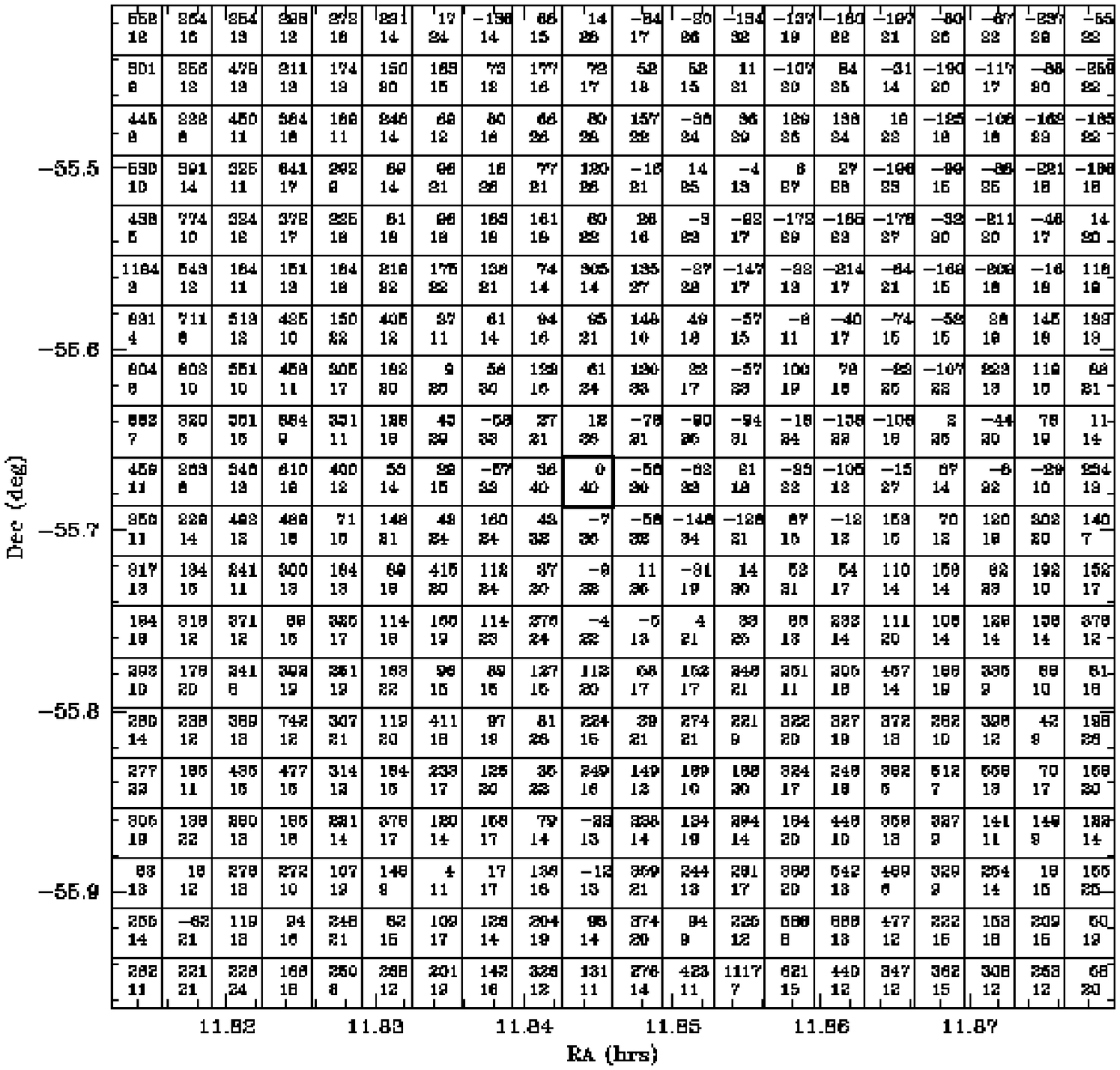,width=11cm,height=11cm}}
\caption{The top number in each  pixel of the grid is the
reddening values 
in milli-magnitudes, calculated with respect
to the fiducial main sequence and corresponding to each subregion, while
 the bottom number in each subregion is the 
 number of stars used to compute the relative reddening in that pixel.
 The  fiducial region is marked by a thick  box. }
\label{rednumseq}
\end{figure*}

As in \citet{vonb01}, we have 
obtained the   reddening map at the cluster distance shown 
in Figure  \ref{redmapseq}, where the darker regions correspond to the higher
relative reddening; for reference, the position of the 
fiducial region and the cluster region are also indicated
by the small box and the circle, respectively. 
The   reddening values in milli-magnitudes,
relative to the  fiducial sequence and
corresponding to each subregion, are the top number in each pixel of the 
grid  shown in Figure  \ref{rednumseq}. The bottom number in each pixel is the 
number of stars used to determine the relative reddening in that pixel. 
The thicker box indicates
the position of the fiducial region. 
We note that higher reddening regions correspond to lower
star numbers with respect to the average star number in the surrounding
regions. Nevertheless, we are aware that
the relative reddening values are significant for the
central regions, where the cluster population is dominant, but are not  
significant for the external
regions of our field of view where the stellar population is dominated by
  field stars; in fact, these stars lay at different distances
 and the fiducial main sequence used to determine the reddening is not 
 representative in these regions.

\begin{figure*}[!ht]
\centerline{\psfig{figure=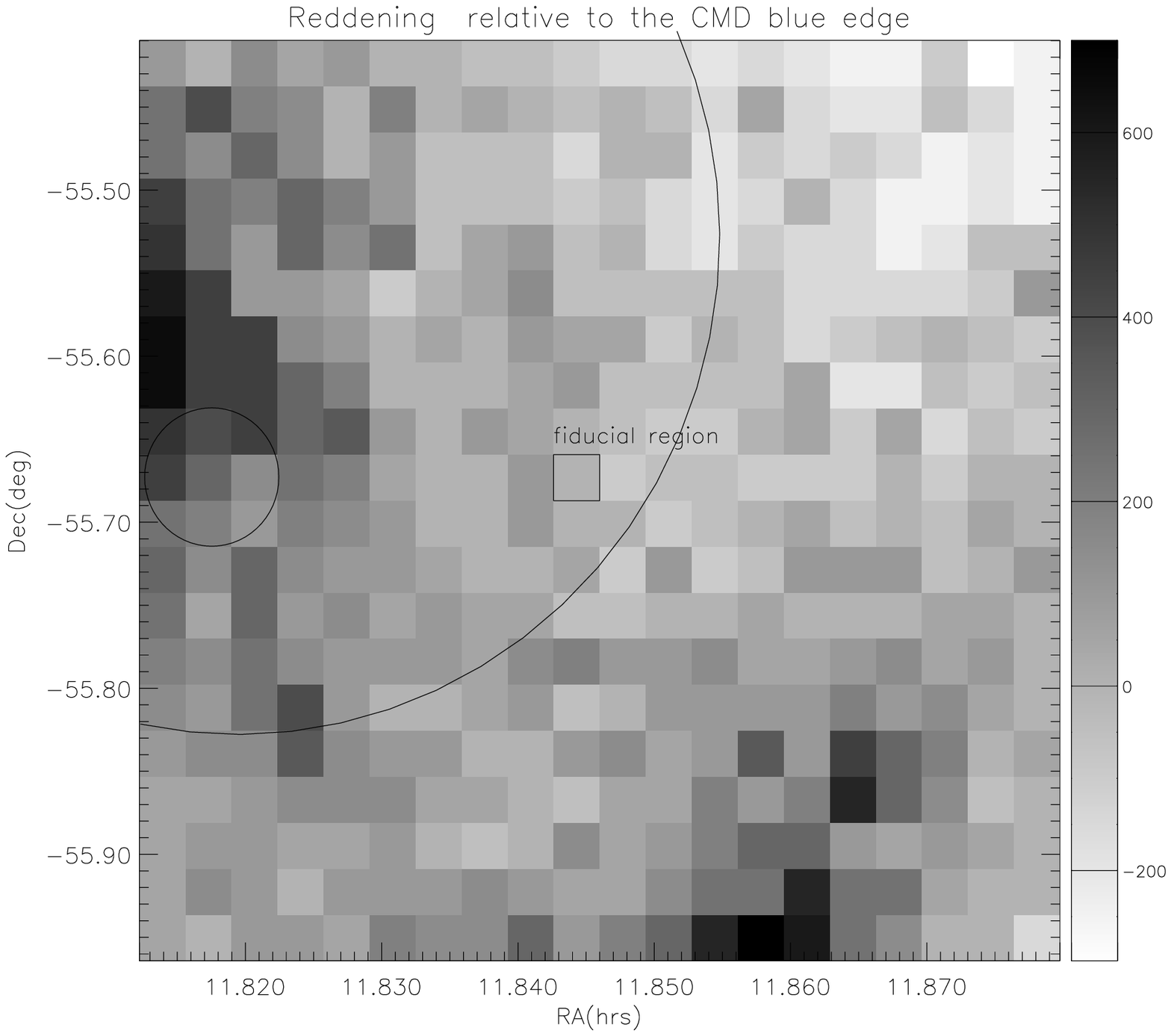,width=11.0cm,height=11.0cm}}
\caption{Greyscale reddening map of the field stars in the \ngc3960 field. 
In this figure,
the darker regions correspond to the higher relative reddening, calculated with
respect to the blue edge of the CMD in the fiducial region. 
The position of the fiducial region
is indicated as the small box. The small circle and the circular line indicate
the position of the dark nebulae included in the Hartley et al. (1986) and
Feitzinger $\&$ Stuewe (1984) catalogues.}
\label{redmapgig}
\end{figure*}

\begin{figure*}[!ht]
\centerline{\psfig{figure=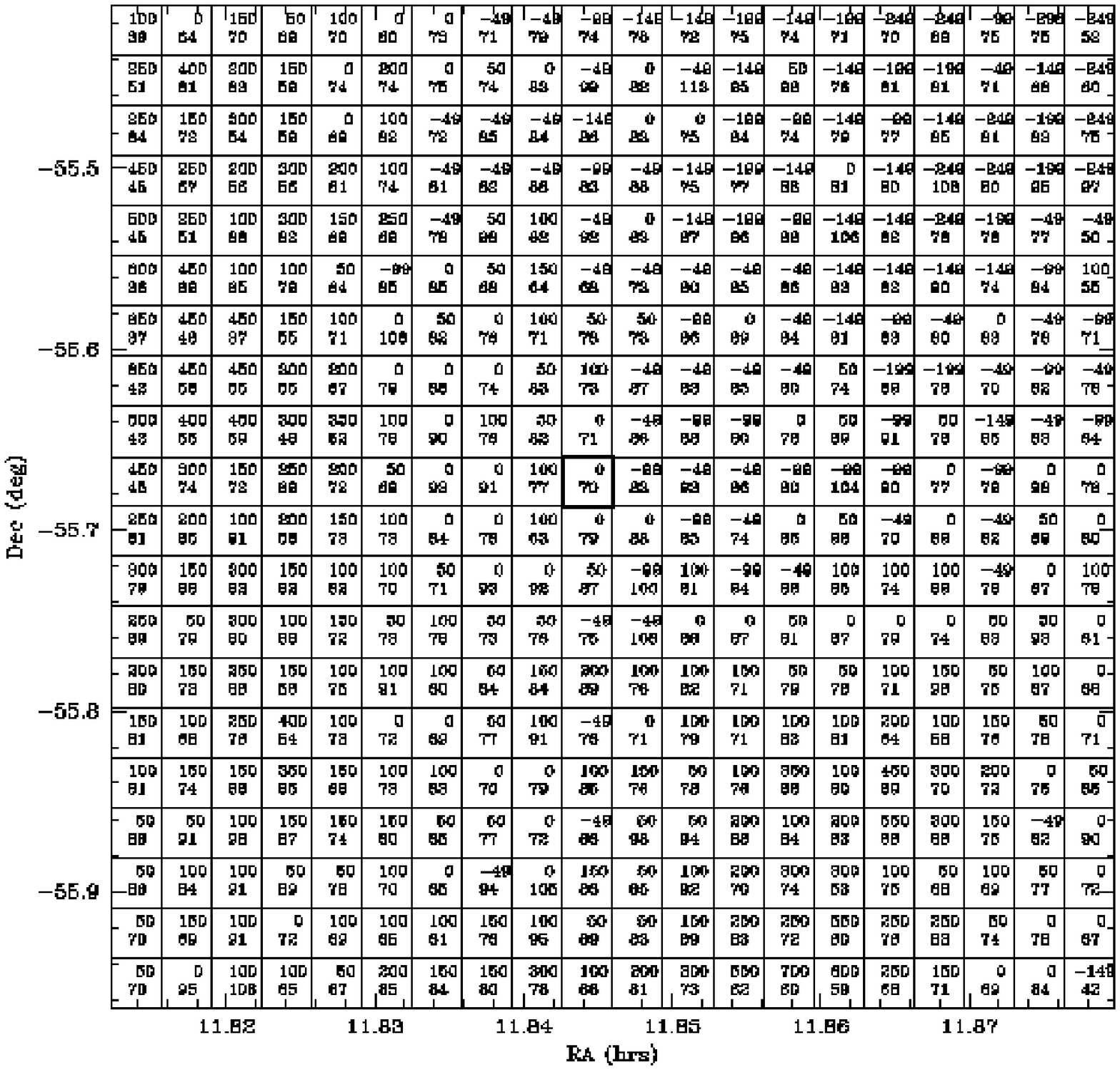,width=11cm,height=11cm}}
\caption{The top number in each  pixel of the grid is the
reddening values 
in milli-magnitudes, calculated with respect
to the  blue edge of the CMD in the fiducial region and corresponding 
to each subregion, while
 the bottom number in each subregion is the 
 number of stars used to compute the relative reddening in that pixel.
 The  fiducial region is marked by a thick  box. }
\label{rednumgig}
\end{figure*}

In order to estimate  the 
differential reddening affecting the field stars, we have calculated
the separation of each star with $18.5<V<22.0$ and $0.0<(V-I)<2.8$, 
from the 
blue edge  of the CMD in the fiducial region \citep{vonh02}. 
The relative reddening affecting these stars has been estimated
as the mode of the $\Delta(V-I)$ distribution. We note that 
we are approximating the relative reddening to a single value for all the 
field stars, although we are aware that the relative reddening is almost
certainly spread out in distance. Using a single reddening value  for each
subregion 
is an approximation, since clearly we cannot determine the reddening for each
star. However, our value can be used to determine an average contamination of 
cluster candidates, 
as will
be discussed in Section  \ref{opticaldata}.
  The obtained results are shown in
Figure  \ref{redmapgig}, where the fiducial region position is also indicated;
the corresponding values in milli-magnitudes and the number of the 
stars used to determine the relative reddening  are reported
  in Figure  \ref{rednumgig}, where the position of the fiducial region is marked 
  by a thicker box.
In Figure  \ref{redmapgig}, the  small circle and the circular line
indicate  the position of
the dark nebulae included in the \citet{hart86}  and  
\citet{feit84} catalogues.
As already mentioned, \citet{dutr02} estimated  $E(B-V)=0.59$
in the region corresponding to the small circle.
 On the other hand, in this region
we have measured a  shift of $\Delta E(V-I)=0.45$   with respect 
to the CMD blue edge in the fiducial region. 
Using the relation given in \citet{muna96}, this
shift corresponds to $\Delta E(B-V)=0.36$. By subtracting this value to the
reddening estimated by \citet{dutr02}, we find, in our fiducial region, 
$E(B-V)=0.23$, that is consistent with the average 
value  $E(B-V)=0.29$ given by \citet{jane81} for the central region of the cluster. 
This external test shows that our differential reddening
estimate is consistent with   previous reddening estimates. 
 
We note that the two reddening maps are qualitatively consistent with each 
other, both showing the two regions with highest reddening, located in 
the north-west
and in the south-east part of the field, and the lowest reddening region,
located in the north-east part of the field.
In the cluster region the median of the differences between the reddening values
derived from the two methods is 
$A_V=0.08$ mag 
with a standard deviation of 0.25 mag.
We have used the   
reddening values  relative to the cluster main sequence 
(Figure  \ref{rednumseq}) 
to  correct  magnitude and color measurements of the stars within  the
{\it cluster region}, where we expect most of the cluster members 
to be located, and the  
reddening values  relative to the blue edge of the CMD to correct magnitudes 
and colors of the stars in the {\it field region},
 where the stellar population is dominated
by field stars.
We find that, within the cluster region, the $E(V-I)$ values range from 0.21 
up to 0.78, assuming that  $E(V-I)=0.36$ ($E(B-V)=0.29$ \citep{jane81})
 in the fiducial region;
corrections in $E(V-I)$   up to 1 mag have been, instead,  applied for the 
most reddened field stars.

The reddening corrected $V$ vs. $V-I$  CMD in the {\it cluster region} 
is shown in Figure  \ref{vicorr} ({\it left panel}).
We note that  the main sequence is better defined with respect to the
 uncorrected  CMD shown in the lower-right panel of Figure   \ref{rawcmd}.
The Turn Off and the RGB stars, which were spread out in Figure   \ref{rawcmd},
are now well traced.  Of course, our procedure computes an average correction
on each pixel of our map and a residual spread in the CMD remains.
\section{Cluster Fundamental Parameters}
\label{param}
Our photometric data together with  very recent stellar models
allow us to determine the cluster parameters, previously 
estimated by \citet{jane81} from a sample of
only 318 stars with $V<16.5$. 
 
In order to find  the age and the distance  of the cluster, we have considered a
set of different  age 
theoretical isochrones recently calculated by
\citet{piet03}, with metallicity $Z=0.01$,
as spectroscopically determined by \citet{frie93}.
By fixing the reddening value $E(B-V)\simeq 0.29$, derived by \citet{jane81}, 
we have vertically shifted the set of reddened isochrones
in the $V$ vs. $(V-I)$
CMD of  the stars  in the {\it cluster region}.
The distance that better fit the upper part of main sequence stars
($15\lesssim V\lesssim 17$) corresponds to a distance modulus of 
$(V-M_V)_0=11.35$. Since we do not know the individual membership 
we can only constrain the age within the range of ages of isochrones
limiting most of the stars close to the Turn Off. With this criterion
the cluster age is between 0.9 and 1.4 Gyr.

\begin{figure*}[!ht]
\centerline{\psfig{figure=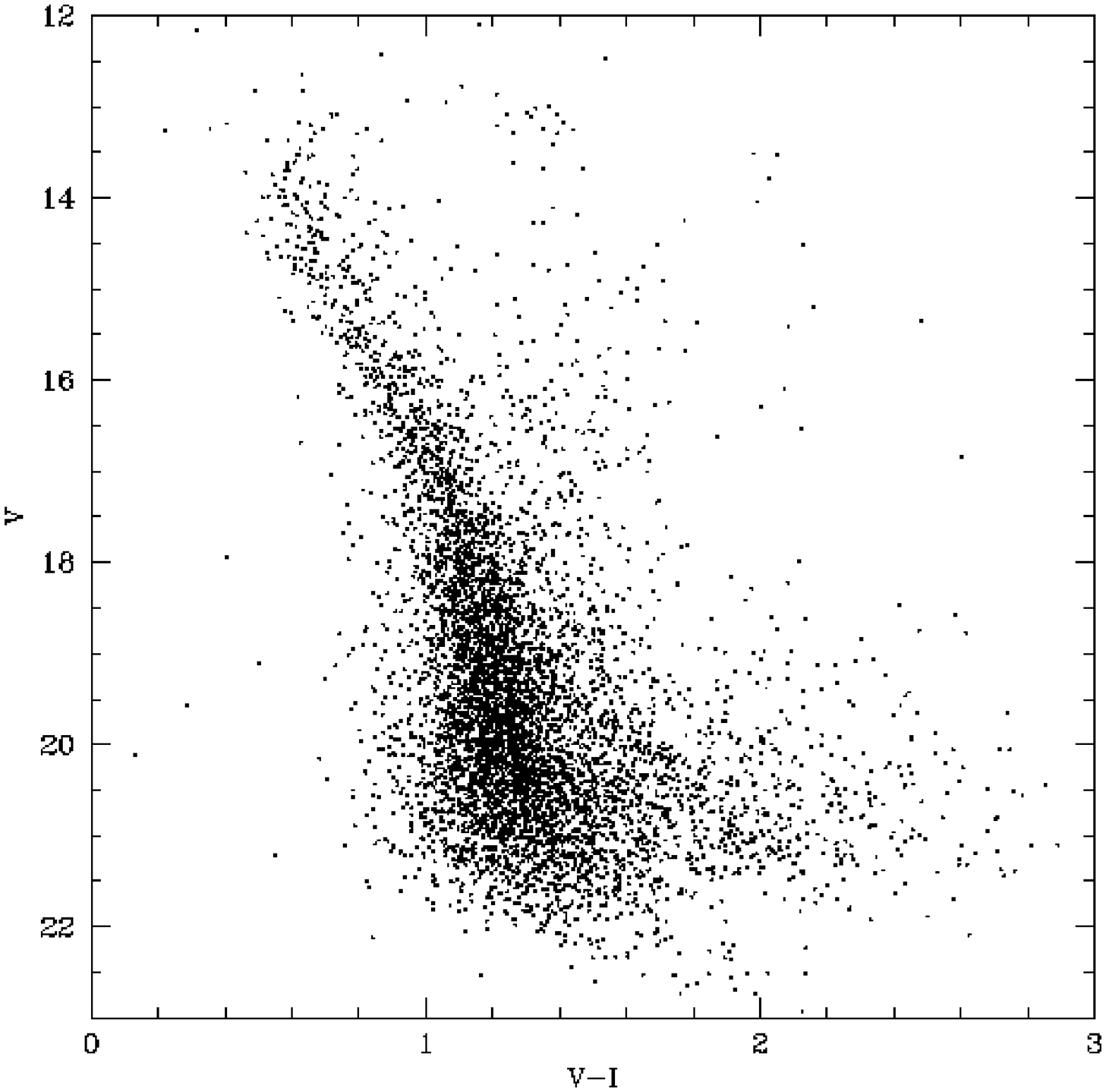,width=8cm,height=10cm} \psfig{figure=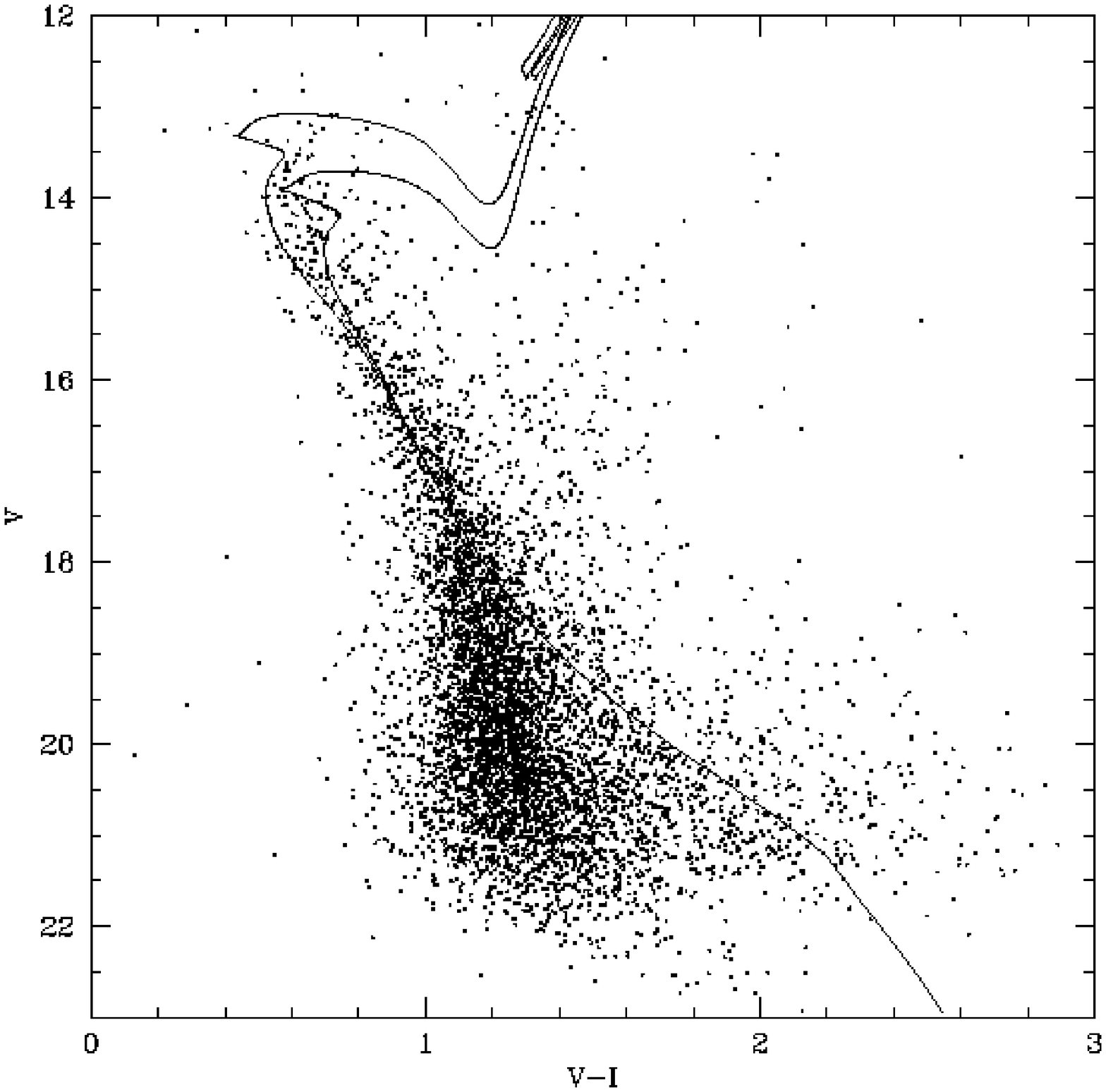,width=8cm,height=10cm}}
\caption{{\it Left panel}: The $V$ vs. $V-I$  diagram of the
stars within the cluster region, corrected for differential reddening.
{\it Right panel}: The  $V$ vs. $V-I$   diagram of the
stars within the cluster region, corrected for differential reddening,
as compared to the Pietrinferni
et al. (2003) isochrones of age   0.9 and 1.4 Gyr,
for the metallicity $Z=0.01$.
Using the color excess of $E(B-V)=0.29$ \citep{jane81},
a distance modulus of $(V-M_V)_0=11.35$ mag has been derived.} 
\label{vicorr}
\end{figure*}

Our results therefore suggest an age for \ngc 3960
older than 
that given in \citet{jane81} and a distance of about 1850 pc, 
that is  slight larger than the value given by the same author 
(d=1660 pc). 
In the right panel of Figure  \ref{vicorr}  we show the  
 $V$ vs. $V-I$  
 diagram  of the stars within  the {\it cluster region} where
the adopted theoretical isochrones have been superimposed.
\section{Empirical Cluster Locus and Photometric Selection of Candidate 
Cluster Members} 
\label{removing}
\subsection{Optical Data}
\label{opticaldata}
As already mentioned, the reddening corrected $V$ vs. $V-I$ CMD computed
from the stars within the cluster region
is  highly contaminated by field stars, as shown by the fact
 that only the bright
cluster main sequence is clearly visible. 
In order to
 define a sample of candidate cluster
members useful to study the cluster population, 
we need to  define photometrically the complete cluster locus in the CMD.
This result  is usually achieved by 
superimposing theoretical stellar models to the
observed CMD where the complete cluster main sequence is clearly
visible, which is not our case.
  To check the agreement of theoretical isochrones,
that suffer from a number of uncertainties \citep{vonh02,groc03},
with our data, 
we have empirically 
recovered the cluster main sequence   
using a statistical  subtraction of the CMD density distribution, 
as described below. A similar approach has been used by \citet{vonh02}.
  
We have used the field region to perform a statistical subtraction of 
the contaminating
field stars in the CMD derived in the cluster region.  The result of 
this  subtraction defines
the locus of the cluster main sequence \citep{chen98,vonh02,baum03}.
 
\begin{figure}[!htb]
\centerline{\psfig{figure=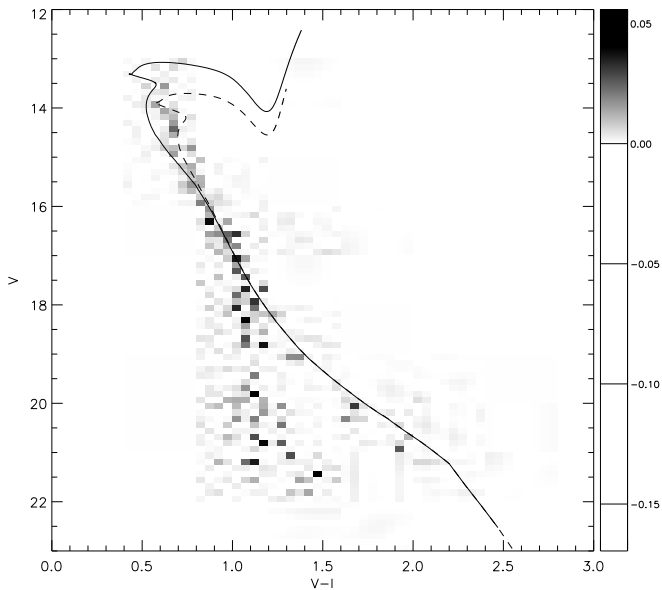,width=9cm,height=9cm}}
\caption{Greyscale CMD density map of the cluster obtained by subtracting the 
$V$ vs. $V-I$  CMD density maps, computed for the "{\it clean field region}"  
and for  the "{\it cluster region}". 
Only boxes with a $1~\sigma$ excess are shown. The solid and dashed lines are,
respectively, 
the 0.9  and 1.4 Gyr isochrones 
with metallicity $Z=0.01$, computed by Pietrinferni et al. (2003), scaled to the 
observed CMD using the cluster parameters given in Section  \ref{param}.}
\label{vivmap}
\end{figure}

Using our reddening corrected photometric data, we have defined  a
 {\it "clean field region"}   by selecting from the 
 {\it "field region"} all those  subregions in Figure  \ref{rednumgig} 
whose 
relative reddening correction was between $-0.150$ and $0.200$, 
corresponding to the range of reddening, estimated for the field stars within the 
{\it "cluster region"}. We have chosen these limits in order to consider only
field  stars affected by an absorption similar to that of the cluster stars. 
 For both the cluster and the field regions, we have constructed
a grid of boxes of  $\Delta V=0.125$
and $ \Delta (V-I)=0.05$
on the $V$ vs. $V-I$ plane. Thus,
we have built a greyscale CMD density map  of the cluster
 by subtracting the unreddened CMD density map of the field stars,
 from the  unreddened CMD density map
 computed for
 the {\it "cluster region"}. 
Figure \ref{vivmap} shows the resulting map where only the boxes showing a
$1~\sigma$ excess are shown.
This excess should follow the cluster main sequence.
The  cluster  main sequence is clearly present
 down to $V\sim18$, while the subtraction is noisier for $V>18.5$,
the region of the cluster main sequence  more strongly contaminated 
by field  stars, 
where the main sequence exhibits a lower star density;
as we have already mentioned, our relative reddening correction suffers from 
the uncertainty due to the spread of the reddening of individual stars, which 
prevents us from obtaining 
  an accurate field star CMD estimate. This explains
the lower star density in the low mass cluster main sequence 
 and the noise of the bluer
field stars. However, in this subtraction the faint cluster population 
($V>18.5$) is well 
separated by the field star population this allowing us to have a well defined
empirical locus of the cluster main sequence.

In Figure  \ref{vivmap}, the 0.9 and 1.4 Gyr isochrones \citep{piet03}
have been superimposed onto the CMD density map 
  using the cluster parameters 
found in Section  \ref{param}. The good agreement between the empirically recovered
cluster main sequence locus and the adopted theoretical model  
down to at least $V\sim18$, confirm both
that the parameters determined in Section  \ref{param} are suitable to the 
cluster and that the adopted theoretical models reproduce the cluster main
sequence over the mass range for which the IMF will be fitted.

We have used both the empirical cluster locus and the mentioned theoretical
isochrones to define a photometric cluster member sample.
To this aim, we have defined a strip in the observed CMD; the lower and upper
limits take into account the spread of the reddening determined  above.
Furthermore, the upper limit of our strip is  
displaced upward
by 0.75 mag in order to include   binary stars.  
 We have defined as candidate cluster members a total of 2119  stars
lying in the cluster region that, within their
photometric errors, belong to this strip. The photometric/astrometric catalogue
of these candidate cluster members is given in Table 5\footnote{available 
in the electronic form at the CDS via anonymous ftp
to {\tt cdsarc.u-strasbg.fr (130.79.128.5)} or via 
{\tt http://cdsweb.u-strasbg.fr/cgi-bin}}, where we report RA and Dec (J2000)
coordinates in decimal degrees, an identification number for each star, the 
$V$, $B$, $I$ magnitudes and their uncertainties.  
\subsection{IR Data}
\label{irdata}
As already stressed in the previous Sections, our field of view is strongly affected
by differential reddening, thus making  uncertain the field star CMD estimate 
and   the study of the cluster population.
To have an external check of our results, we have used 
near-infrared 
$JHK_s$ data from 2MASS survey of the same field of view \citep{carp01}, 
since IR bands are not as much sensitive to reddening effects as optical bands. 
To have a sample as reliable as possible,
only stars detected in all the $JHK_S$ bands, with
magnitudes measured from point spread-function fitting or aperture photometry 
have been selected from the 2MASS catalogue.

\begin{figure}[!ht]
\centerline{\psfig{figure=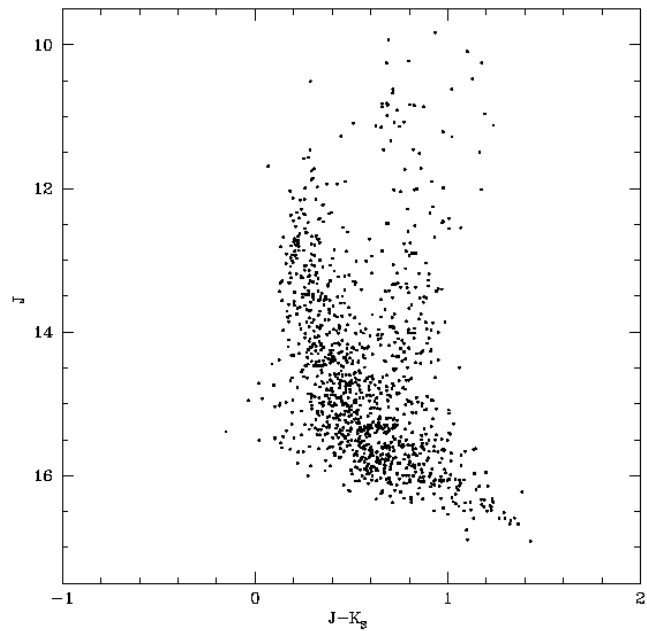,width=9cm,height=9cm}}
\caption{$J$ vs. $J-K_S$ CMD  obtained from the 2MASS catalogue of the stars 
detected 
within 7 arcmin from the cluster centroid. The diagram shows only
stars revealed in all  the $JHK_S$ bands with magnitudes  measured from
point spread-function fitting or aperture photometry.}
\label{jkj}
\end{figure}

The first step has been to construct IR CMDs  of the stars in the
cluster region. Figure  \ref{jkj} shows the $J$ vs. $J-K_S$ CMD
of 1162 stars detected in this area; we can see that the cluster main 
sequence is 
revealed from  higher star density sequence down to   $J\simeq 16.5$,
 although it remains
rather broad probably due to a  large scatter in the colors of the low 
Signal-to Noise ratio sources. 

\begin{figure}[!htb]
\centerline{\psfig{figure=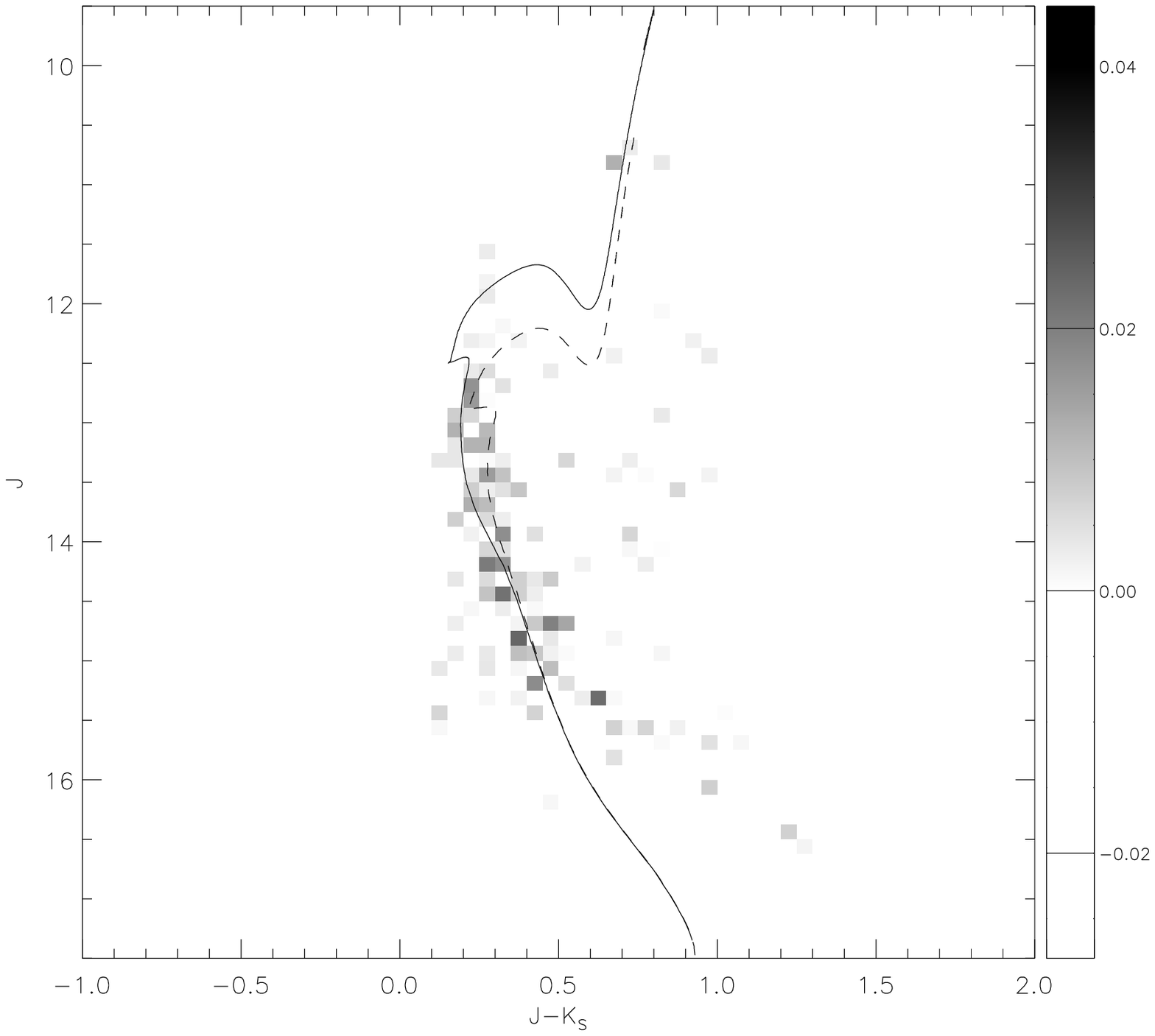,width=9cm,height=9cm}}
\caption{Greyscale density map of the cluster obtained by subtracting the 
$J$ vs. $J-K_S$ CMD density map, computed for the "{\it field region}" 
(cf Section  \ref{irdata}), 
from the $J$ vs. $J-K_S$ CMD density map, 
computed for the "{\it cluster region}". Only
boxes with a  $1~\sigma$ excess are shown. The solid and dashed lines are,
respectively,
the 0.9  and 1.4 Gyr isochrones 
with metallicity $Z=0.01$, computed by \citet{piet03}, scaled to the 
observed CMD using the cluster parameters given in Section  \ref{param}.}
\label{jkjmap}
\end{figure}

In this diagram, the clump of stars at $J\simeq11$ and
$(J-K_S)\simeq0.7$ are certainly Red Giant Branch stars of \ngc 3960,
while the lower density stars in the almost vertical sequence 
are mainly giant field stars. 
As already done with optical data, we have built the greyscale density map  
of the cluster with IR data
 by subtracting the field  CMD density map from the $J$ vs. $J-K_S$ CMD 
 density map computed for
 the {\it "cluster region"}. In order to estimate an 
optical-independent field star distribution, 
we have defined 
 as {\it "field region"}  the farther annulus area with radius $R$  between
 30' and 45' from the cluster centroid. 
Figure \ref{jkjmap} shows the IR resulting map where only the boxes having a
$1~\sigma$ excess are shown. We note as   the cluster main 
sequence is well defined at least  down to $J=15.8$, that is the 2MASS 
completeness and reliability magnitude limit.  
In order to verify  the cluster parameters, 
we have superimposed in Figure  \ref{jkjmap} 
the 0.9 and 1.4 Gyr isochrones of \citet{piet03}, adopted also in the optical case,
using the parameters that we have determined in Section  
\ref{param} and  the reddening law derived by \citet{card89}.
 We again find a good
agreement between theoretical models and observational data down to 
$J\simeq 15.5$,
while for fainter stars, the model do not reproduce the empirical cluster main
sequence; this can be due both to the strong uncertainties affecting IR 
color-temperature relations \citep{groc03} and to the  
photometric errors in this magnitude range. 
However, the agreement of the adopted theoretical model
with the bright cluster main sequence
confirms again the reliability of the cluster parameters found 
in Section  \ref{param}.

Once the empirical cluster main sequence has been recovered from the IR data, 
we have applied, as done with the optical data, a photometric selection of 
candidate cluster members. Using  the adopted theoretical isochrone for the
upper main sequence and a fiducial line for the lower one, we have defined
a strip in the CMD, including the boxes about the main sequence
with a $1~\sigma$ excess;
in this way we have selected a sample of 749  IR cluster candidates  
within 
the {\it  "cluster region"} and belonging, 
according to their photometric errors, to this strip. 
The photometric/astrometric catalogue
of these candidate cluster members is given in Table 6
\footnote{available in electronic form at the CDS via anonymous ftp
to {\tt cdsarc.u-strasbg.fr (130.79.128.5)} or via 
{\tt http://cdsweb.u-strasbg.fr}}, where we report RA and Dec (J2000)
coordinates in decimal degrees, an identification number for each star, the 
$V$, $B$, $I$ magnitudes and their uncertainties. 
\section{Luminosity and Mass Functions}
\label{lfmf}
One of the main objectives of this study is to derive the stellar mass function
down to the limiting magnitude reached in this survey, that is $V\simeq 22$. 

In Section  \ref{opticaldata} we have found a sample of optical candidate cluster
members, but  
such  photometrically selected sample contains the contribution of
foreground and background sources that has to be subtracted in order to 
estimate the
number of stars belonging to \ngc3960.    Since neither spectroscopic
observations, nor proper motions are available for this cluster, 
we have used the following  statistical approach to estimate
the number of the cluster members.

\begin{figure}[!htb]
\centerline{\psfig{figure=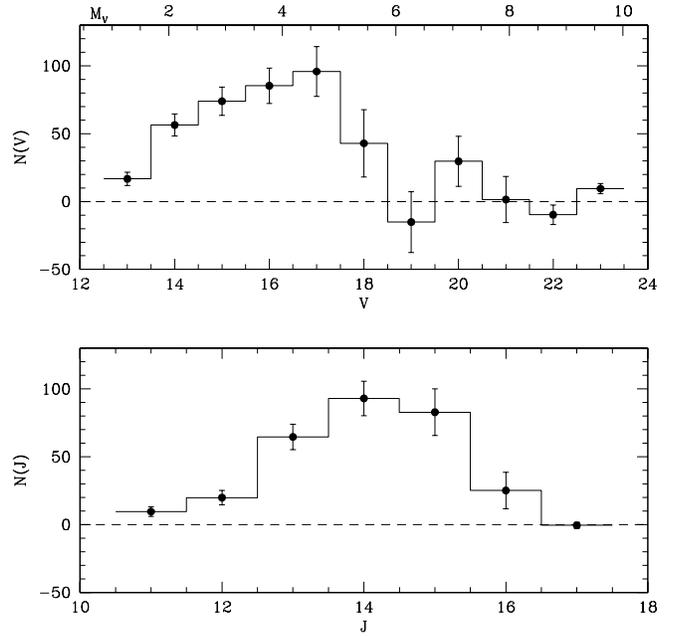,width=9cm,height=9cm}}
\caption{Luminosity functions of \ngc3960, corrected for the
 field star contribution,
obtained from optical data ({\it top panel}) and from IR data 
({\it bottom panel}). The dashed lines indicate the zero value of the
distributions.}
\label{lfvj}
\end{figure}

First, we have computed the total distribution of the optical candidate 
members lying 
in the "{\it cluster region}", as a
function of the $V$ magnitude; second, we have computed the $V$ distribution of
the candidate members falling in the "{\it clean field region}" 
defined in Section  \ref{opticaldata}, unreddened relatively to the fiducial region;
 then, this field distribution has been normalized
to the  "{\it cluster region}" area and then subtracted to the total one. The resulting
$V$ luminosity function of the cluster is shown in the top panel of 
Figure  \ref{lfvj}, where the  values of the absolute magnitude ($M_V$),
corresponding to the visual magnitudes are also indicated in the top axis. 
We note that for $V>18.5$, the distribution is in some case consistent with the zero
value and in one case, is even negative(!). This result is consistent with the   
low star density found in Figure  \ref{vivmap} in the same magnitude range. 
As already mentioned, this finding is very likely due to the critical field
star distribution estimate about $V=19.5$, where the contribution of the field
 stars is dominant. For $V>20$, the deficiency of stars is probably due to
incompleteness of our catalogue that, based on the artificial star tests,
is $100\%$ complete down to $V\simeq 20$. In addition,  due to dynamical
evolution, a percentage
of such faint cluster
 stars is expected to be found also outside our estimated 
{\it cluster region}, that is where we have estimated the field star 
distribution,
and this makes it difficult to have a correct estimate of the cluster luminosity
function in this magnitude range.  

An independent IR luminosity distribution of the cluster has been obtained  
using the same statistical approach and the sample of IR candidate cluster members
obtained in Section  \ref{irdata}. As for the optical case, the contribution of field
stars has been computed using the "{\it field region}"  defined in Section  
\ref{irdata}. The resulting J luminosity function is shown in the bottom panel of 
Figure  \ref{lfvj}. Also in this case, the result is reliable for $J<15.8$, 
 that is in 
the magnitude range where the 2MASS catalogue is complete. This limiting magnitude
corresponds to $V\sim 17$, for a main sequence star at the cluster distance.

\begin{figure}[!htb]
\centerline{\psfig{figure=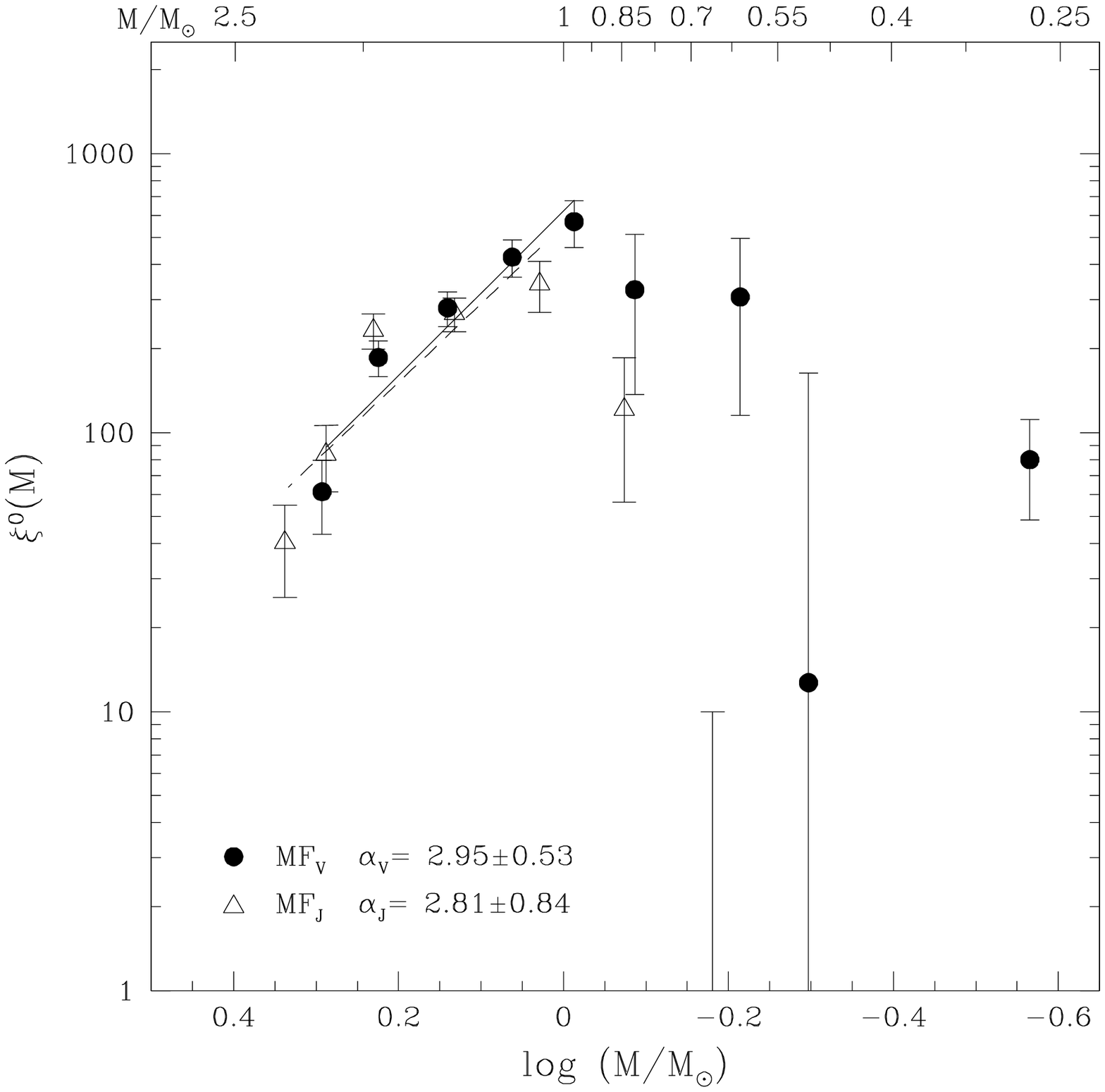,width=9cm,height=9cm}}
\caption{Comparison of the \ngc3960 mass function obtained from the optical data
({\rm filled points}) and from IR data ({\rm empty triangles}).  The 
solid line is the
power law fit to the mass function obtained from the $V$ luminosity function,
while  the dashed line is the power law fit to the mass function obtained from
the $J$ luminosity function. $\xi(M)$ values are given in number per logarithmic
mass unit. The slopes of the power law fitting with the rms residuals of the least
squares fit, as obtained from the two distributions, are also indicated.}  
\label{mfcomp}
\end{figure}

In order to compare these results, obtained from two independent surveys,
we have transformed the $V$ and $J$ luminosity functions into the corresponding
mass functions. We have used
in both cases, the Mass-Luminosity Relation (MLR) computed by \citet{piet03}. 
The derived mass functions are shown 
in Figure  \ref{mfcomp}. As already stressed, the results are  reliable for
$M>1M_\odot$, where both  surveys are complete. For this mass range, we have
computed a linear fit to both  distributions and  have found that 
the two functions have a slope of $2.95 \pm 0.53$ and $2.81\pm0.84$, for
 the $V$ and $J$ luminosity functions, respectively. 

The agreement of the mass function obtained from the optical data with the one
obtained from IR data indicates that the applied corrections for differential
reddening and, therefore, the field star distribution estimate, 
are  reliable. 
Nevertheless, the very large errors corresponding to the optical mass function
in the range  $M<1M_\odot$ are the result of
the already mentioned problems in this mass
range that prevent from having a reliable  estimate of the 
low mass function of \ngc3960.

\section{Summary and Conclusions}
\label{summary}
This Chapter presents  photometry and astrometry  of the stars  
in the field of the open cluster \ngc3960, falling in the $34\times 33$ arcmin
square field of the WFI camera of the MPG/ESO 2.2\,m  Telescope. Our survey 
reaches a limiting magnitude $V\sim 22$; based on  artificial star
tests,  it is $100\%$ complete 
 down to $V\sim 20$, where the photometric accuracy is better than $6\%$.  
 Comparison of the positions of the 
 stars in our survey with those  of the GSC\,2.2,
 used as reference catalogue, indicates that the final astrometric accuracy 
is $<$0\Sec2. 

The photometric data have been used to construct the  $V$ vs. $B-V$ and the
 $V$ vs. $V-I$  CMDs, where the high contribution of the disk population  
allows us to identify only the cluster  bright star main sequence, 
that  appears rather broadened due to strong  differential
reddening. A detailed spatial study of the  $V$ vs. $V-I$ CMD allows us to
evaluate the relative reddening of  20 subregions, each of about  
1\Min7$\times$1\Min7,
 with respect to a fiducial region corresponding to the
cluster centroid. The relative reddening has been computed 
from the cluster main sequence, for the stars falling in the {\it cluster region},
and using the blue edge of the CMD  for  the field stars, which,
being at distances different from that of the cluster, are affected by   
different reddening.   Our results indicate that, 
within the cluster region, the $E(V-I)$ values range from 0.21 
up to 0.78;
corrections in $E(V-I)$   up to 1 mag have been, instead,  applied for the 
most reddened field stars. 

The final reddening map traces regions of low star density, 
that are consistent with two areas, indicated in literature 
as {\it dark nebulae}, that are regions of the sky where the apparent
surface density of stars is reduced compared to surrounding regions.
Comparison of our reddening estimate in the smallest
of  these
{\it dark nebulae}, with the one reported in literature, 
allows us to estimate a color excess $E(B-V)$ of about 0.23 in the region
corresponding to the cluster centroid. This value is very similar to the value
0.29 given  by \citet{jane81} for the cluster. 

Using the reddening corrected CMD  and recent stellar models, 
we have found that the cluster has an age between 0.9 and 1.4 Gyr, its distance 
modulus
is $(V-M_V)_0=11.35$ and it is located about 1850 pc  from the Sun.

In order to minimize the field star contamination in the CMD,  a
statistical subtraction of the CMD density map  has been performed 
to empirically recover the whole cluster main sequence. Using the 
resulting density map, we have been able to compare 
the empirical cluster main sequence with the adopted theoretical
models, thus allowing us to verify the reliability of our estimated cluster
parameters. Furthermore, this empirical cluster main sequence has been used
to identify a strip in the optical CMD where cluster stars are expected to be
located;   2119 candidate cluster
members fall in this strip.

In order to have an external check of our results, we have built an analogous
CMD density map using IR data from the 2MASS catalogue.
Taking advantage of the availability of the entire sky coverage of the 2MASS
catalogue and in order to avoid the above mentioned dark nebulae,
a different 
control region for field stars,
located farther than the one used in the optical case, has been used. 
The obtained empirical
cluster main sequence in the $J$  vs.  $J-K_S$ plane has been compared to  
theoretical models using the cluster parameters estimated in this work. 
  The result of this comparison 
shows, again,  the correctness of the estimated cluster parameters.
As in the optical case, the IR empirical main sequence has been used to define a
strip in the $J$ vs. $J-K_S$ CMD 
and therefore a sample of IR candidate cluster members.

From  the two samples of candidate cluster members, obtained from the optical and
IR surveys, and using the respective control field, the luminosity distributions 
for the cluster, as
a function of the $V$ and $J$ magnitudes, have been computed. Using the same MLR,
the two distributions have been transformed into two independent mass function
determinations of the cluster.
In the mass range where the two data set are complete ($M<1M_\odot$), 
the two mass functions have  been fitted to a power law having indices 
$\alpha_V=2.95\pm0.53$ and $\alpha_J=2.81\pm0.84$ in $V$ and in $J$,
    respectively, while the
   Salpeter mass function in this notation has index $\alpha=2.35$.
Our values are both  consistent  with the data presented in the 
$\alpha$ vs. log $M$ plot of \citet{krou02} for other open clusters.

The good agreement of the two distributions, furthermore, 
suggests both that the reddening
correction of our catalogue and the estimates for the field star contamination
are reliable and correct. However, 
the low mass range is affected both by completeness problems as well as by
likely mass segregation effects
preventing a reliable derivation of the cluster luminosity and mass functions 
for $M<1M_\odot$.

A future spectroscopic study of this region is planned in order to 
measure the radial velocities crucial to have an independent membership criterion
and to determine individual reddening of each star. 
\begin{acknowledgements}
We thank S. Cassisi and A. Pietrinferni for providing their most recent models and
M. Zoccali and E. Flaccomio for useful suggestions that greatly improved the
presentation of this paper. This work has been partially supported by MIUR.
\end{acknowledgements}
\bibliographystyle{aa}
\bibliography{n3960}

\begin{thebibliography}{29}
\expandafter\ifx\csname natexlab\endcsname\relax\def\natexlab#1{#1}\fi

\bibitem[{{Baume} {et~al.}(2003){Baume}, {V{\' a}zquez}, {Carraro},
  {et~al.}}]{baum03}
{Baume}, G., {V{\' a}zquez}, R.~A., {Carraro}, G., {et~al.} 2003, \aap, 402,
  549

\bibitem[{{Cardelli} {et~al.}(1989){Cardelli}, {Clayton}, \& {Mathis}}]{card89}
{Cardelli}, J.~A., {Clayton}, G.~C., \& {Mathis}, J.~S. 1989, \apj, 345, 245

\bibitem[{{Carpenter}(2001)}]{carp01}
{Carpenter}, J.~M. 2001, \aj, 121, 2851

\bibitem[{{Chen} {et~al.}(1998){Chen}, {Carraro}, {Torra}, {et~al.}}]{chen98}
{Chen}, B., {Carraro}, G., {Torra}, J., {et~al.} 1998, \aap, 331, 916

\bibitem[{{de La Fuente Marcos}(1997)}]{fuen97}
{de La Fuente Marcos}, R. 1997, \aap, 322, 764

\bibitem[{{Dutra} \& {Bica}(2002)}]{dutr02}
{Dutra}, C.~M. \& {Bica}, E. 2002, \aap, 383, 631

\bibitem[{{Feitzinger} \& {Stuewe}(1984)}]{feit84}
{Feitzinger}, J.~V. \& {Stuewe}, J.~A. 1984, \aaps, 58, 365

\bibitem[{{Friel} \& {Janes}(1993)}]{frie93}
{Friel}, E.~D. \& {Janes}, K.~A. 1993, \aap, 267, 75

\bibitem[{{Grocholski} \& {Sarajedini}(2003)}]{groc03}
{Grocholski}, A. \& {Sarajedini}, A. 2003, \mnras, in press

\bibitem[{{Hartley} {et~al.}(1986){Hartley}, {Tritton}, {Manchester},
  {et~al.}}]{hart86}
{Hartley}, M., {Tritton}, S.~B., {Manchester}, R.~N., {et~al.} 1986, \aaps, 63,
  27

\bibitem[{{Janes}(1977)}]{jane77}
{Janes}, K.~A. 1977, \pasp, 89, 576

\bibitem[{{Janes}(1981)}]{jane81}
{Janes}, K.~A. 1981, \aj, 86, 1210

\bibitem[{{Kroupa}(2002)}]{krou02}
{Kroupa}, P. 2002, Science, 295, 82

\bibitem[{{Landolt}(1992)}]{land92}
{Landolt}, A.~U. 1992, \aj, 104, 340

\bibitem[{{Miller}(1972)}]{mill72}
{Miller}, E.~W. 1972, \aj, 77, 216

\bibitem[{{Munari} \& {Carraro}(1996)}]{muna96}
{Munari}, U. \& {Carraro}, G. 1996, \aap, 314, 108

\bibitem[{{Pietrinferni} {et~al.}(2003){Pietrinferni}, {Cassisi}, {Salaris},
  {et~al.}}]{piet03}
{Pietrinferni}, A., {Cassisi}, S., {Salaris}, A., {et~al.} 2003, \aap, in
  preparation

\bibitem[{{Piotto} {et~al.}(1999){Piotto}, {Zoccali}, {King}, {Djorgovski},
  {Sosin}, {Rich}, \& {Meylan}}]{piot99a}
{Piotto}, G., {Zoccali}, M., {King}, I.~R., {et~al.} 1999, \aj, 118, 1727

\bibitem[{{Prisinzano} {et~al.}(2003){Prisinzano}, {Micela}, {Sciortino},
  {et~al.}}]{pris03}
{Prisinzano}, L., {Micela}, G., {Sciortino}, S., {et~al.} 2003, \aap, 404, 927

\bibitem[{{Schlegel} {et~al.}(1998){Schlegel}, {Finkbeiner}, \&
  {Davis}}]{schl98}
{Schlegel}, D.~J., {Finkbeiner}, D.~P., \& {Davis}, M. 1998, \apj, 500, 525

\bibitem[{{Stetson}(1987)}]{stet87}
{Stetson}, P.~B. 1987, \pasp, 99, 191

\bibitem[{{Stetson}(1990)}]{stet90}
{Stetson}, P.~B. 1990, \pasp, 102, 932

\bibitem[{{Stetson}(1994)}]{stet94}
{Stetson}, P.~B. 1994, \pasp, 106, 250

\bibitem[{{Stetson}(2000)}]{stet00}
{Stetson}, P.~B. 2000, \pasp, 112, 925

\bibitem[{{Stetson} \& {Harris}(1988)}]{stet88}
{Stetson}, P.~B. \& {Harris}, W.~E. 1988, \aj, 96, 909

\bibitem[{{van den Bergh} \& {McClure}(1980)}]{berg80}
{van den Bergh}, S. \& {McClure}, R.~D. 1980, \aap, 88, 360

\bibitem[{{von Braun} \& {Mateo}(2001)}]{vonb01}
{von Braun}, K. \& {Mateo}, M. 2001, \aj, 121, 1522

\bibitem[{{von Hippel} {et~al.}(2002){von Hippel}, {Steinhauer}, {Sarajedini},
  {et~al.}}]{vonh02}
{von Hippel}, T., {Steinhauer}, A., {Sarajedini}, A., {et~al.} 2002, \aj, 124,
  1555

\bibitem[{{Zoccali} {et~al.}(2003){Zoccali}, {Renzini}, {Ortolani},
  {et~al.}}]{zocc03}
{Zoccali}, M., {Renzini}, A., {Ortolani}, S., {et~al.} 2003, \aap, 399, 931

\end{thebibliography}

\end{document}